\documentclass[prd,eqsecnum,showpacs,aps,nofootinbib,onecolumn,preprintnumbers,floats,superscriptaddress,floatfix,amsmath,amssymb]{revtex4}
\usepackage{graphicx}

\usepackage{exscale}

\usepackage{watermark}

\renewcommand{\vec}[1]{\mathbf{#1}}

\begin{document}


\bibliographystyle{apsrev}




\title{The missing link: a nonlinear post-Friedmann framework for  small and large scales}



\author{Irene Milillo}\email{irene.milillo@gmail.com}
\affiliation{Institute of Cosmology and Gravitation, University
of Portsmouth, Dennis Sciama Building, Burnaby Road, Portsmouth PO1 3FX, United Kingdom}
\affiliation{Dipartimento di Fisica, Universit\`a di Roma ``Tor Vergata'',
via della Ricerca Scientifica 1, 00133 Roma, Italy}
\affiliation{
Enea - Centro Ricerche Casaccia
Via Anguillarese, 301 - 00123 ROMA, Italy}

\author{Daniele Bertacca}\email{daniele.bertacca@gmail.com}
\affiliation{Argelander-
Institut f\"ur Astronomie, Auf dem H\"ugel 71, D-53121 Bonn, Germany}
\affiliation{Physics Department, University of the Western Cape, Cape Town 7535, South Africa }
\affiliation{Dipartimento di Fisica ``Galileo Galilei", Universit\'a di Padova, via F. Marzolo 8-35131, Italy}
\affiliation{Institute of Cosmology and Gravitation, University
of Portsmouth, Dennis Sciama Building, Burnaby Road, Portsmouth PO1 3FX, United Kingdom}

\author{Marco Bruni}\email{Marco.Bruni@port.ac.uk}
\affiliation{Institute of Cosmology and Gravitation, University
of Portsmouth, Dennis Sciama Building, Burnaby Road, Portsmouth PO1 3FX, United Kingdom}

\author{Andrea Maselli}\email{andrea.maselli@roma1.infn.it}
\affiliation{Dipartimento di Fisica, Sapienza Universit\`a di Roma \& Sezione INFN Roma1, Piazzale Aldo Moro 5, 00185, Roma, Italy}
\affiliation{Center for Relativistic Astrophysics, School of Physics, Georgia Institute of Technology, Atlanta, Georgia 30332, USA.}

\date{\today}
\begin{abstract}
We present a  nonlinear post-Friedmann framework for structure formation, generalizing to cosmology the weak-field (post-Minkowskian) approximation,  unifying the treatment of small and large scales.  We consider a universe filled with a pressureless fluid and a cosmological constant $\Lambda$, the theory of gravity is Einstein's general relativity and the background is the standard flat $\Lambda$CDM cosmological model.
 We expand the metric and the energy-momentum tensor in powers of $1/c$,  keeping the matter density and  peculiar velocity   as  exact fundamental variables.  We assume the Poisson gauge, including scalar and tensor modes  up to $1/c^4$ order and vector modes up to $1/c^5$ terms.
Through a redefinition of the scalar potentials as a
resummation of the metric contributions at different orders, we obtain a complete set of nonlinear
equations, providing a unified framework to study structure formation from small to superhorizon scales, from the  nonlinear Newtonian to the linear relativistic regime. We explicitly show the validity of our scheme in 
 the two limits: at leading order we recover the fully nonlinear equations of Newtonian cosmology; when linearized, our equations become those for scalar and vector modes of first-order relativistic perturbation theory in the Poisson gauge.  Tensor modes are nondynamical at the $1/c^4$  order we consider (gravitational waves  only appear at higher order): they are purely nonlinear and describe a distortion of the spatial slices  determined at this order by a  constraint, quadratic in the scalar and vector variables. 
 
 The main results of our analysis are as follows: (a) at leading order a purely 
 Newtonian nonlinear energy current  sources a frame-dragging gravitomagnetic 
 vector potential, and (b)  in the leading-order Newtonian regime and in the linear relativistic regime the two scalar metric potentials are the same, 
while the nonlinearity of general relativity makes them different.

Possible applications of our formalism  include the calculations of the vector potential \cite{Bruni:2013mua,2015arXiv150100799T} and the  difference between the two scalar potentials from Newtonian N-body simulations, and the extension of Newtonian approximations used in structure formation studies,  to include  relativistic effects.

\end{abstract}


\maketitle

\section{Introduction}

The $\Lambda$CDM model \cite{Peebles:1984ge,Efstathiou:1990xe}  has emerged in the last few decades as the standard ``concordance" model of cosmology \cite{Tegmark:2003ud}. Beyond photons and baryons, the main components of $\Lambda$CDM are cold dark matter (CDM), able to cluster and  form structures,  and the cosmological constant $\Lambda$,  responsible for the observed acceleration  of the Universe's expansion. 

$\Lambda$CDM is based on Einstein general relativity (GR), and on  the
cosmological principle, i.e.\ a cosmological version of the  Copernican principle \cite{ellis2012relativistic,Maartens:2011yx}. 
This request for the Universe to be, {\it on average}, homogeneous and isotropic translates, in the language
of spacetime, into assuming a Robertson-Walker metric. With this, Einstein equations
reduce to the Friedmann equations; the solutions of these equations
are the Friedmann-Lemaitre-Robertson-Walker (FLRW) models. 
Cosmic microwave background (CMB) anisotropy measurements \cite{deBernardis:2000gy} have established that the Universe is, to a great degree, spatially flat, as confirmed by  recent Planck \cite{Ade:2013zuv} and baryon acoustic oscillations (BAO) data \cite{Aubourg:2014yra}. In this standard scenario, small primordial inflationary perturbations on top of the FLRW background grow and produce the CMB fluctuations and the large-scale structure that we observe at low redshift. 

The  theoretical tools that we use to study the growth of the  large-scale structure   are basically two:  {\it i)} relativistic perturbation theory \cite{Bertschinger:1993xt,Malik:2008im} is used to describe fluctuations in the early Universe,  in the CMB, and 
in the matter  density field on very large scales; {\it ii)}
Newtonian methods, notably N-body simulations \cite{Bertschinger:1998tv}, are used to study the  growth of structures in the nonlinear regime, at late times and small scales.
 Lagrangian perturbation theory (LPT) \cite{Ehlers:1996wg,Bernardeau:2001qr} (i.e.\  the Zel'dovich approximation  or its second-order extension, 2LPT)  
is typically used to  set up initial conditions for N-body simulations; LPT and other Newtonian approximations are also used  to model   nonlinear scales, e.g.\ BAO and CDM halos \cite{Crocce:2007dt,Matsubara:2008wx,Taruya:2010mx,Wang:2014gda,McCullagh:2014jsa,Seljak:2015rea}.

Observational cosmology has now reached an unprecedented precision, allowing stringent tests on models of the Universe. The tightest constraints come from the CMB \cite{Planck:2015xua}.  A number of probes exist at low redshift, such as supernovae \cite{Betoule:2014frx},   the local measurements of the expansion rate \cite{Riess:2011yx} and  the three-dimensional mapping of the  large-scale structure \cite{SDSS,WiggleZ}. In particular, rich clusters, lensing and redshift space distortion allow the measurement of the growth of clustering \cite{Song:2008qt,Samushia:2011cs}.
The standard $\Lambda$CDM model is very well supported by all these observations. 
However,
a tension emerges  in the framework of the base $\Lambda$CDM when  
parameter values  measured from 
low-redshift probes are compared with the values obtained from the CMB,
 as recently pointed out
\citep{Macaulay:2013swa,Verde:2013wza,Battye:2014qga,Ruiz:2014hma}.
In particular,  this tension  shows up in 
 a different growth of clustering in LSS at different epochs \cite{Ade:2013zuv}, as confirmed by the Planck release:  
``as in the 2013 analysis, the amplitude of the fluctuation spectrum is found to be higher than inferred from some analyses of rich cluster counts and weak gravitational lensing'' (see  \cite{Planck:2015xua} and Refs.\ therein). Measurements of redshift space distortion also suggest less clustering than the CMB \cite{Macaulay:2013swa,Battye:2014qga}. Notably, adding a  variable total neutrinos mass as extra parameter to the base $\Lambda$CDM model relieves the tension \cite{Battye:2014qga,Verde:2013wza}. 

Given that the presence of a CDM component is widely accepted, while a  cosmological constant  has its own problems \cite{Weinberg:1988cp}, much work is gone into exploring alternatives to $\Lambda$. 
A number of possibilities have been proposed which, 
in essence, can be divided in three groups. One  is to keep GR and modify the energy-momentum content, the dark sector in particular, replacing $\Lambda$ with some form of dark energy \cite{amendola2010dark}.
A second alternative that has been widely explored in the last decade is to replace GR with a modified theory of gravity \cite{Nojiri:2006ri,Clifton:2011jh}. A third, more radical,  option is that of abandoning the cosmological principle \cite{Clarkson:2010uz}, considering inhomogeneous models and the possibility that the observed acceleration of the Universe is the result of backreaction \cite{Buchert:2011sx,Clarkson:2011zq}, either dynamical or optical \cite{Kolb:2009rp}. 

Galaxy surveys  are now aiming at the same 1\% precision as  CMB measurements: for instance, the detection of BAO by the Sloan Digital Sky Survey has recently allowed the first  1\% level cosmological constraint  by a galaxy survey \cite{Anderson:2013zyy}. 
 In addition,  future surveys such as Euclid \cite{Laureijs:2011gra,Amendola:2012ys,Scaramella:2015rra} and The Square Kilometer Array \cite{Jarvis:2015tqa,Schwarz:2015pqa,Kitching:2015fra} will reach scales of the order of the Hubble horizon. 
It is, therefore, crucial that the theory used to make predictions and to interpret observations is developed with a matching accuracy. 
In particular, while exploring alternatives to the standard $\Lambda$CDM scenario is an interesting challenge,
it seems timely to refine the  theoretical modelling of $\Lambda$CDM, bridging the above mentioned gap between  the Newtonian treatment of  nonlinear small scales and the relativistic  description of large scales. Ultimately, going beyond the Newtonian approximation in simulations of large-scale structure should be important in order to take into account causal, retardation and other GR effects that may be non-negligible for simulations that aim at $\sim1\%$ accuracy, in view of future surveys such as Euclid \cite{Laureijs:2011gra}, 
 on scales of the order of the Hubble horizon.

A first step in this direction would be to include GR corrections in the initial conditions for simulations, using a dictionary based on first-order perturbation theory  \cite{Chisari:2011iq}, cf.\ \cite{Green:2011wc}; also, first-order GR effects on horizon scales  must be taken into account in interpreting bias and non-Gaussianity \cite{Bruni:2011ta}. However, while the Poisson equation in Newtonian gravity establishes a linear relation between the gravitational potential and the matter density field, the intrinsic nonlinearity of GR unavoidably generates new effects, even when perturbations are small. For instance,  an initially Gaussian curvature inflationary perturbation translates into an effective  non-Gaussianity of the density field in the matter era \cite{Bar05,Bar10,Bruni:2013qta,Bruni:2014xma}, a nonlinear effect that should be taken into account in initial conditions for simulations. But the Poisson equation  embodies action-at-the-distance; i.e., it is the mathematical representation of the acausal nature of Newtonian gravity. Thus the ultimate step forward would be that of 
 investigating the effects of  relativistic  nonlinearity in cosmological structure formation,
including GR corrections in the evolution of the matter density field in N-body simulations, as well as in approximate treatments of the nonlinear regime \cite{Chisari:2011iq,Green:2011wc,Bar05,Bar10,Bruni:2011ta,Bruni:2013mua,Kopp:2013tqa,Bruni:2013qta,Bruni:2014xma,Adamek:2014xba,Green:2014aga,2014CQGra..31w4005V,Rigopoulos:2014rqa,2015arXiv150100799T,Yoo:2014kpa}.

Aim of this paper is to present a new  nonlinear relativistic post-Friedmann (PF) formalism which, in essence, 
is a generalization to cosmology of the post-Minkowski (weak-field) approximation, married with the fundamental assumption of the post-Newtonian (PN) approximation that velocities are small. 
Our goal is a relativistic framework valid on all scales, and including the full nonlinearity of Newtonian gravity at small scales. In this framework - assuming a flat $\Lambda$CDM background and a fluid description of matter -  the exact equations of Newtonian cosmology \cite{Peebles:1980za,Peacockbook} appear as a consistent approximation of  the full set of Einstein equations, determining leading-order terms (which we call 0PF)  in the metric.  GR corrections (which we call 1PF) appear next, and are of two types: terms quadratic in the Newtonian variables (we may call these  proper PN terms), and proper linear GR terms. Once linearised, the nonlinear approximate equations we obtain for a set of appropriately resummed variables are  those for the scalar and vector sectors of first-order relativistic perturbation theory \cite{Ma:1995ey}. Tensor modes are purely nonlinear and nondynamical; i.e., they do not describe propagating gravitational waves, but rather a distortion of the spatial slices. 

Let us clarify how our PF formalism differs from  the traditional PN approach \cite{Weinberg:1972zz} and its various applications in cosmology.
In a contemporary perspective \cite{2014grav.book.....P}, the correct derivation of the post-Newtonian approximation on the flat background spacetime  follows consistently from the post-Minkowski approximation once the $v/c\ll 1$ assumption is made, which also implies to neglect time derivatives with respect to space derivatives\footnote{The contemporary derivation of the post-Newtonian approximation \cite{2014grav.book.....P} should be contrasted with the traditional implementation (see e.g.\ \cite{Weinberg:1972zz}) which suffers from various ambiguities and inconsistencies \cite{PhysRevD.28.2363,2014grav.book.....P}. By and large,  cosmological applications of the post-Newtonian approximation follow the traditional approach (see below).}.  However, in cosmology the background is a FLRW solution, in our case the flat $\Lambda$CDM model, and what we can assume to be small are peculiar velocities, not the change with time of the physical distance between two arbitrary observers. To illustrate the point, adopting a Newtonian perspective, 
in the absolute space of Newtonian cosmology one uses the background comoving coordinates $\vec{x}$ as the Eulerian grid, and the physical position of a fluid element is $\vec{r}=a\vec{x}$. Then $\dot{\vec{r}}=H\vec{r}+\vec{v}$, where $\vec{v}=a\dot{\vec{x}}$ is the physical (or proper) peculiar velocity \cite{Bertschinger:1993xt},  representing the deviation from the Hubble flow; $\dot{\vec{x}}$ is the peculiar velocity with respect to the comoving grid \cite{Peacockbook}.  
Then, should we assume that $|\dot{\vec{r}}|\ll c$, we would end up with an approximation only valid at small scales, well below the Hubble horizon, $|\vec{r}|\ll c H^{-1}$. 
In addition, traditionally the PN formalism has been developed to study GR corrections to the orbits of isolated objects. Thus, the focus is on the equations of motion, rather than on a consistent approximation of the full set of Einstein equations. In developing his PN  treatment of relativistic hydrodynamics, having in mind applications to relativistic stars, Chandrasekhar \cite{Chandrasekhar:1965ch} also focused on the equations of motion for the fluid, as derived from the conservation equations. In both cases, the metric and other variables are expanded\footnote{Formally, the expansion is with respect to 
a small dimensionless parameter representing the ratio $v/c$; by dimensional analysis  the ratio of the gravitational field and $c^2$ is also dimensionless and can be expressed in terms of the same parameter. In this way the parameter, which in practice  is  the inverse of the speed of light $1/c$, indicates the relativistic weight of each term in the expansion \cite{stachel}.}  in powers of $1/c$, 
then the expansion is applied to the equations of motion. As it is well known, a correction to the time-time component of the Minkowski metric is all is needed to obtain the Newtonian equations of motion. All other corrections are then considered post-Newtonian. 
We will discuss this point in detail in Sec. \ref{pas-act}, where, 
 following \cite{Misner:1973mi}, we call this approach focused on the equations of motion "passive". 
 In cosmology, however, we find desirable to  consider an "active" approach, where the space-space and time-time components of the metric are equally weighted, leading directly  to a consistent treatment, order by order in the expansion parameter $1/c$, of the full set of Einstein equations\footnote{The approximate conservation equations then consistently follow from the contracted Bianchi identities, as in the exact theory.}. In view of applications, a useful byproduct of this active approach is that the equal weighing of the different metric components is what is needed to obtain the correct photon trajectories, i.e.\ the null geodesics that are at the base of the causal structure of the spacetime. 

Various early and more recent works have applied the traditional  PN expansion in cosmology: in  \cite{PhysRevLett.61.2175, Tomita:1988to, Tomita:1991to, PhysRevD.53.681} the PN equation of motion for particles in the expanding Universe are derived; in \cite{Kofman:1994kf} the authors clarify the  role of the electric and magnetic part of the Weyl tensor in the Newtonian approximation; in \cite{Takada:1997bk} and \cite{Matarrese:1995sb} the PN analysis is given in Lagrangian coordinates using, respectively, the 3+1 and 1+3 framework; in \cite{Hwang:2005mg} the authors derive a complete set of field and hydrodynamic equations in Newtonian-like forms.  Other remarkable works following a PN method are those of Szekeres \cite{Szekeres:1999ny, Szekeres:2000ki}. Finally,   \cite{Carbone:2004iv} follows   a "hybrid approximation scheme", somehow closer in spirit to our PF approach,  a mix  between standard cosmological perturbation theory and PN approximation, also obtaining equations for the generation of gravitational waves.

All these  works  apply the standard iterative approach of the PN expansion. Here instead we focus on defining a set of resummed 1PF variables, and on deriving the set of nonlinear evolution and constraint equations they satisfy.
Specifically, assuming the  Poisson (or conformal-Newtonian \cite{Ma:1995ey,Bertschinger:1993xt,poisgaug,Malik:2008im}) gauge, we expand the metric  in powers of $1/c$ and  in the equations we retain all scalar terms up to order $1/c^{4}$ (and all vector terms up to $1/c^{5}$),   instead of peeling-off the different orders, cf.\ \cite{Szekeres:2000ki,Szekeres:1999ny}. This  leads to  a final set of equations that differs from those of the  works mentioned above. Partly following the PN tradition,  the 0PF and 1PF orders, respectively  refer to terms $1/c^{2}$ and $1/c^{4}$, relevant for scalar potentials. The  0PF terms are Newtonian, the 1PF terms contain the GR corrections.
Finally, we stress that, in order to obtain a system of equations  with a well-posed Cauchy problem, we should  consider terms of the next 2PF order, i.e.\ $1/c^6$ \cite{Szekeres:2000ki,Szekeres:1999ny}. 

In summary, the main goal of this paper is to present a set of nonlinear resummed equations up to 1PF order which retain in full the nonlinearity of Newtonian theory on small scales and all linear relativistic perturbation theory on large scales. The 1PF scheme is, therefore, capable of describing, in a unified framework and at the relevant leading orders, the evolution of large-scale structure   on all  scales of cosmological interest. 
Our two  main results simply follow from the  analysis of our nonlinear equations: at leading order a purely 
 Newtonian nonlinear energy current  sources a frame-dragging gravitomagnetic 
 vector potential;  in the leading-order Newtonian regime and in the linear relativistic regime the two scalar metric potentials are the same, 
while the nonlinear 1PF equations for the resummed scalar potentials imply that the nonlinearity of GR makes them different.

The paper is organized as follows. After  defining the various metric  terms  in Sec. \ref{PNmetric},  we obtain the stress energy tensor in Sec. \ref{matter-variables},  the field equations in Sec. \ref{EinsteinEq} and mass and momentum conservation   equations in Section \ref{ConservationEq}. We then consider our equations in two opposite limits: the Newtonian  regime on small scales, neglecting  $O(1/c^4)$ terms (see  Sec. \ref{Newtonianregime}) and the linear regime 
on large scales through the linearization of the equations (see Sec. \ref{linear_limit}). In Sec. \ref{NonlinearPN}, we first define suitable resummed variables, then we obtain a consistent set of nonlinear equations describing their evolution.
In Sec. \ref{Conclusions}, we draw our main conclusions. 
Finally, in  Appendix \ref{Appendix},  we  apply the PF expansion to  the Riemann and Ricci tensors.

\section{Newtonian and post-Friedmann metric variables}\label{PNmetric}

We consider a homogeneous and isotropic FLRW flat background where two kinds of perturbation terms are added to the metric,  representing two different levels of accuracy. The first will give the Newtonian regime, i.e.\ an approximate solution of Einstein equations such that the dynamics is described by the exact nonlinear equations of Newtonian cosmology for pressureless matter.   The second will give the  relativistic corrections, adding terms to the Newtonian equations.
  In our post-Friedmann framework the goal is to calculate these relativistic terms; in this spirit, we shall refer to the   Newtonian approximation  as the 0PF (or leading) order, and  to the first   relativistic corrections as  1PF order.

The expanding parameter for the inhomogeneous perturbations is formally given by $1/c$, 
so that  the components of the metric tensor in the line element
\begin{equation}
-c^2d\tau^2=d s^2=g_{\mu\nu}dx^\mu dx^\nu
\end{equation}
 can be written  as
\begin{subequations}
\label{metric}
\begin{eqnarray}
\label{g_00}
g_{00}&=&-\left[1-\frac{2U_N}{c^2}+\frac{1}{c^4}(2U_N^2-4U_P)\right]+O\left(\frac{1}{c^6}\right)\text{,}\\
\label{g_0i}
g_{0i}&=&-\frac{a}{c^3}B^N_i-\frac{a}{c^5}B^P_i+O\left(\frac{1}{c^7}\right)\text{,}\\
\label{g_ij}
g_{ij}&=&a^2\left[\left(1+\frac{2V_N}{c^2}+\frac{1}{c^4}(2V_N^2+4V_P) \right)\delta_{ij}+\frac{1}{c^4}h_{ij}\right]+O\left(\frac{1}{c^6}\right)\text{,}
\end{eqnarray}
\end{subequations}
where  $\tau$ is  proper time  and $a(t)$ is the scale factor of the FLRW background. Note that we assume that both the scale factor and the metric are dimensionless, while  the coordinates have dimension of a length.
Greek indices take the values $0,1,2,3$ and refer to spacetime coordinates,  Latin indices  refer to the spatial coordinates. In particular in our post-Friedmann scheme, for a proper ``powers of  $c$" order counting, it is important to note that    the time coordinate is  $x^0=ct$. Here the Kronecker $\delta_{ij}$ represents the metric on the  flat spatial  slices of the background; the spatial Cartesian coordinates are  understood as an Eulerian  system of reference, cf.\ \cite{2014PhRvD..89f3509R,2014PhRvD..90l3503R}.

Our  metric is a generalization  to cosmology of Chandrasekhar's  metric  for post-Newtonian hydrodynamics \cite{Chandrasekhar:1965ch} (cf.\  \cite{Hwang:2005mg} for a cosmological application). However, it is important to remark the difference of our post-Friedmann approach from the standard post-Newtonian one \cite{lrr-2006-4,test,Weinberg:1972zz,2014grav.book.....P}. In the latter, the focus is on the equation of motion for matter, hence only the leading order $g_{00}$ metric perturbation is Newtonian, while the $g_{ij}$ is post-Newtonian. In our approach the focus will instead be on the complete set of Einstein equations. We aim at a self-consistent approximate set of equations at each order, similarly to the post-Minkowski (weak field) approximation \cite{mit-notes,2014grav.book.....P}. Then, consistency of the  Einstein equations dictate that the $g_{ij}$ and $g_{00}$ metric perturbations must be of the same order (see Sec. \ref{Newtonianregime} for the Newtonian approximation). 
 Anticipating these results, the indices $N$ and $P$ label Newtonian and post-Friedmann quantities. The $N$ quantities are the only relevant one at the 0PF order of approximation of Einstein equations, i.e. in the Newtonian regime. The $P$ quantities appear at  the 1PF order.

We assume the Poisson  (or conformal Newtonian) gauge \cite{Ma:1995ey,Bertschinger:1993xt,poisgaug,Malik:2008im,Uggla:2012gg}, so that the three-vectors $B^N_i$ and $B^P_i$ are divergenceless, $B_i^{N,i}=0$ and $B_i^{P,i}=0$, and  $h_{ij}$ is transverse and tracefree (TT); i.e., it represents pure tensor modes $h^i_{\;i}=h_{ij}^{\;\;\;,i}= 0$.   Commas in front of indices have the standard meaning of partial derivatives.

 Having completely fixed the gauge leaves us with six degrees of freedom at each order: the two scalars $U_N$ and $V_N$ at leading order, and $U_P$  and $V_P$ at 1PF order; the two independent components of  $B^N_i$ ($B^P_i$ at 1PF);  the two independent components of  $h_{ij}$. However, the latter only appears in the equations at 1PF order. At leading order, Einstein equations  impose $U_N=V_N$, i.e.\ a single scalar gravitational potential in the Newtonian regime, as expected. At leading order,  $B^N_i$ is determined by the (vector part of the) Newtonian  energy current, and cannot be set to zero. It is not dynamical; i.e,\ it doesn't contribute to matter motion. However it appears in the metric and does  affect null geodesics and observables. It can, therefore, be extracted from Newtonian N-body simulations \cite{Bruni:2013mua,2015arXiv150100799T} and it contributes to lensing \cite{Thomas:2014aga}. 

Note that the 1PF corrections consist of quadratic combinations of Newtonian quantities and intrinsic 1PF variables. We have also included tensorial TT modes but, at  1PF  order, they cannot be interpreted as gravitational waves;   they satisfy a constraint equation rather than an evolution equation (see Sec. \ref{sec-TT}, c.f.\ \cite{2014PhRvD..90l3503R}).

\section{Matter variables}\label{matter-variables}

Having defined the metric variables and their weight with respect to the expansion parameter $c^{-1}$, we now look at the matter quantities.
The dimensionless 4-velocity is defined as
\begin{equation}
u^{\mu}:=\frac{dx^{\mu}}{cd\tau}
\end{equation}
 and it  satisfies the usual  relation
\begin{equation}
\label{unit}
g_{\mu\nu}u^{\mu}u^{\nu}=-1\text{,}
\end{equation}
i.e.\ is a unitary timelike vector field. In Newtonian cosmology one uses the background comoving coordinates $\vec{x}$ as the Eulerian grid,  the physical position of a fluid element is $\vec{r}=a\vec{x}$, and  $\vec{v}=a\dot{\vec{x}}$ and $\dot{\vec{x}}$ respectively are is the physical (or proper) peculiar velocity \cite{Bertschinger:1993xt}  (the deviation from the Hubble flow) and the peculiar velocity with respect to the comoving grid \cite{Peacockbook}.  
With this in mind,  it is then natural to define the physical peculiar velocity  as $v^i := a dx^i/dt$.  Therefore, we also define $v_i:=\delta_{ij}v^j$.

Then,
\begin{eqnarray}
\label{ui}
u^i=\frac{dx^i}{cd\tau}=\frac{dx^i}{cdt}\frac{dt}{d\tau}=\frac{v^i}{ca}u^0\text{,}
\end{eqnarray}
and,  using  (\ref{unit}) and keeping terms up to order $c^{-4}$, the 4-velocity components  are 
\begin{subequations}
\label{vel4}
\begin{eqnarray}
&&u^i=\frac{1}{c}\frac{v^i}{a}u^0 \text{,}\\
&&u^0=1+\frac{1}{c^2}\left(U_N+\frac{1}{2}v^2\right)+\frac{1}{c^4}\left[\frac{1}{2}U_N^2+2U_P +v^2V_N+\frac{3}{2}v^2U_N+\frac{3}{8}v^4-B^N_iv^i\right] \text{,}\\
&&u_i= \frac{av_i}{c}+\frac{a}{c^3}\left[-B^N_i+v_iU_N+2v_iV_N+\frac{1}{2}v_iv^2\right]\text{,}\\
&&u_0=-1+\frac{1}{c^2}\left(U_N-\frac{1}{2}v^2\right)+\frac{1}{c^4}\left[2U_P -\frac{1}{2} U_N^2-\frac{1}{2}v^2U_N-v^2V_N-\frac{3}{8}v^4\right] \text{.}
\end{eqnarray}
\end{subequations}

We consider a Universe filled by cold dark matter (CDM), described by a single pressureless (dust) component with energy-momentum tensor
\begin{equation}
T^{\mu}_{\;\;\nu}=c^2\rho u^{\mu}u_{\nu}\text{,}
\end{equation}
where $\rho$ is the mass density.

In the Poisson gauge the components and trace of $T^{\mu}_{\;\;\nu}$ then are
\begin{subequations}\begin{eqnarray}
\label{as}
T^0_{\;\;0}&=&-c^2\rho-\rho v^2-\frac{1}{c^2}\rho\left[2(U_N+V_N)v^2-B^N_iv^i+v^4\right]\;,\\
T^0_{\;\;i}&=&c\rho av_i+\frac{1}{c}\rho a\left\{v_i[v^2+2(U_N+V_N)]-B^N_i\right\}\;,\\
T^i_{\;\;0}&=& -c \frac{1}{a}\rho v^i-\frac{1}{c} \frac{1}{a}\rho v^2v^i\;,\\
T^i_{\;\;j}&=&\rho v^iv_j+\frac{1}{c^2}\rho\left\{v^iv_j[v^2+2(U_N+V_N)]-v^iB^N_j\right\}\;,\\
\label{ad}
T^\mu_{\;\;\mu}&=&T=-\rho c^2 \,.
\end{eqnarray}
\end{subequations}
All these quantities are written in order to explicitly show the different contributions in powers of $c$, where in each expression the first term beyond the background term (if present)  represents the leading Newtonian order, the second the first 1PF correction, etc. Note that there is no approximation in the trace: indeed, the mass density $\rho$ plays the role of a fundamental exact quantity that is not expanded into contributions at different orders. In the following it will be useful to use the density contrast $\delta$, defined as usual: $\delta:=({\rho-\bar{\rho}})/{\bar{\rho}}$, where $\bar{\rho}$ denotes the background matter density. 

\section{Einstein equations}\label{EinsteinEq}
\subsection{Expansion of Einstein equations at 1PF order}
We now consider Einstein field equations for the metric (\ref{metric}), including the cosmological constant $\Lambda$:
\begin{equation}
G^{\mu}_{\;\;\nu}=R^{\mu}_{\;\;\nu}-\frac{1}{2}R \delta^{\mu}_{\;\;\nu}=\frac{8\pi G}{c^4}T^{\mu}_{\;\;\nu}-\Lambda \delta^{\mu}_{\;\;\nu}\;.
\end{equation}
Expanding in powers of $1/c$, in all equations we retain the first two terms of the expansion. We then obtain the following equations, where the dot denotes partial differentiation with respect to coordinate time $t$.\\

\noindent
Time-time component 
\begin{eqnarray}
\label{g00}
G_{\;\;0}^0+\Lambda=\frac{8\pi G}{c^4}T^0_{\;\;0}&\rightarrow&\;  \frac{1}{c^2}\left[3\left(\frac{\dot a}{a}\right)^2-2\frac{\nabla^2 V_N}{a^2}\right]+\frac{1}{c^4}\left[6\frac{\dot a}{a}\dot V_N+6\left(\frac{\dot a}{a}\right)^2U_N-\frac{4}{a^2}\nabla^2V_P+\frac{2}{a^2}\nabla^2V_N^{\;2}-\frac{5}{a^2}V_N^{\phantom{N},i}V_{N,i}\right]\nonumber\\
&&=\frac{1}{c^2}8\pi G\rho+\frac{1}{c^4}8\pi G\rho v^2+\Lambda \;.\nonumber\\
\end{eqnarray}
Spatial component 
\begin{eqnarray}
\label{gij}
G^j_{\;\;i}+\Lambda \delta^j_{\;\;i}=\frac{8\pi G}{c^4}T^i_{\;\;j}&\rightarrow&\;\frac{1}{c^2}\left\{\frac{1}{a^2}(V_N-U_N)_{\;\;,i}^{,j}+\delta_{\;\;i}^j\left[\left(\frac{\dot a}{a}\right)^2+2\frac{\ddot a}{a}-\frac{1}{a^2}\nabla^2(V_N-U_N)\right]\right\}+\frac{1}{c^4}\left\{-\frac{\dot a}{a^2}\left(B_{i}^{N,j}+B^{Nj}_{\phantom{NJ},i}\right)\right.\nonumber\\&&\left.-\frac{1}{2a}\left(\dot B^{Nj}_{\phantom{Nj},i}+\dot B_{i}^{N,j}\right)-\frac{2}{a^2}U_{P,i}^{\phantom{P,i},j}+\frac{2}{a^2}V_{P,i}^{\phantom{P,i},j} +\frac{1}{a^2}U_{N,i}U_N^{\phantom{N},j}-\frac{1}{a^2}V_{N,i}V_N^{\phantom{N},j}
+\frac{1}{a^2}\left(U_{N,i}V_N^{\phantom{N},j}\right.\right.\nonumber\\&&\left.\left.+U_N^{\phantom{N},j}V_{N,i}\right)-\frac{2}{a^2}V_N(V_N-U_N)_{\;\;,i}^{,j}+\delta_{\;\;i}^j\left[2\frac{\dot a}{a}\dot U_N+4\frac{\ddot a}{a}U_N+2\left(\frac{\dot a}{a}\right)^2U_N+ 6\frac{\dot a}{a}\dot V_N+2 \ddot V_N\right.\right.\nonumber\\&&\left.\left.+\frac{2}{a^2}\nabla^2U_P -\frac{2}{a^2}\nabla^2 V_P- \frac{1}{a^2}U_{N,k}U_N^{\phantom{N},k}+\frac{2}{a^2}V_N\nabla^2(V_N-U_N) \right]+\frac{1}{2a^2}\nabla^2h_{\;\;i}^j\right\}\nonumber\\&&=\Lambda\delta_{\;\;i}^{j}-\frac{8\pi G}{c^4}\rho v_i v^j\;.
\end{eqnarray}
Time-space component 
\begin{eqnarray}
\label{gi0}
G^0_{\;\;i}=\frac{8\pi G}{c^4}T^0_{\;\;i}&\rightarrow&\;\frac{1}{c^3}\left[-\frac{1}{2a}\nabla^2B^N_i+2\frac{\dot a}{a} U_{N,i}+2\dot V_{N,i}\right]+\frac{1}{c^5}\bigg[-\frac{1}{2a}\nabla^2B^P_i+
4\frac{\dot a}{a} U_{P,i}+4\dot V_{P,i}+2\dot V_N U_{N,i} +4\frac{\dot a}{a} U_NU_{N,i}\nonumber\\&&+4\dot V_{N,i}V_N+\frac{1}{2a}B^N_{i\phantom{N},k}(V_N-U_N)^{,k}-\frac{1}{2a}B^{N}_{k\phantom{N},i}(U_N+V_N)^{,k}+\frac{1}{a}\nabla^2B^{N}_{i}(V_N-U_N)+\frac{1}{2a}B^N_{i}\nabla^2 V_N\nonumber\\&&+\frac{1}{a}B^{Nk}V_{N,ki}\bigg]=\frac{8\pi G}{c^3}\rho a v_i+\frac{8\pi G}{c^5}\rho a\left\{v_i\left[v^2+2(U_N+V_N)]-B^N_i\right\}\;.\right.
\end{eqnarray}
Finally, neglecting in Eqs.\ (\ref{g00}) and (\ref{gij}) all terms representing inhomogeneities, we obtain the   background equations:
\begin{subequations}
\label{friedmann}
\begin{eqnarray}
\label{friedmann1}
\frac{1}{c^2}\left(\frac{\dot a}{a}\right)^2=\frac{1}{c^2}\frac{8\pi G}{3}\bar\rho+\frac{\Lambda}{3}\;,\\
\label{friedmann2}
\frac{1}{c^2}\left[\left(\frac{\dot a}{a}\right)^2+2\frac{\ddot a}{a}\right]=\Lambda \;.
\end{eqnarray}
\end{subequations}
These are recast into the standard
Friedmann and Raychaudhuri equations for the flat FLRW background
\begin{subequations}
\label{Friedmann}
\begin{eqnarray}
\label{Fried}
\frac{1}{c^2}H^2=\frac{1}{c^2}\frac{8\pi G}{3}\bar\rho+\frac{\Lambda}{3}\;,\\
\label{Ray}
\frac{1}{c^2}\left[\dot{H}+H^2\right]=-\frac{1}{c^2}\frac{4\pi G}{3}\bar\rho +\frac{\Lambda}{3} \;,
\end{eqnarray}
\end{subequations}
after substituting the Hubble expansion scalar $H={\dot{a}}/{a}$.
\subsection{1PF equations for the inhomogeneities}
We now subtract the background parts (\ref{friedmann}) from Eqs.\ (\ref{g00}) and (\ref{gij}), in order to   obtain  equations for the  inhomogeneous quantities. The time-time component of the field equations then gives a generalized Poisson equation:
\begin{eqnarray}
\label{ps}
&-&\frac{1}{c^2}\frac{1}{3a^2}\nabla^2 V_N +\frac{1}{c^4}\left[\frac{\dot a}{a}\dot V_N+\left(\frac{\dot a}{a}\right)^2U_N+\frac{1}{3a^2}\nabla^2 V_N^{\;2}-\frac{5}{6a^2}V_{N,i}V_{N}^{\phantom{N},i}-\frac{2}{3a^2}\nabla^2V_P\right]\nonumber\\&=&\frac{1}{c^2}\frac{4\pi G}{3}\bar\rho \delta +\frac{1}{c^4}\frac{4\pi G}{3}\bar\rho(1+\delta) v^2\;.
\end{eqnarray}
Note that the cosmological constant  disappears from the equations above; it  only directly contributes to the background dynamics, i.e.\  Eqs. (\ref{Friedmann}). Thus perturbations are only affected by $\Lambda$  through their coupling with the Hubble (background)  expansion.

The trace of the space-space component (\ref{gij}) gives\\
\begin{eqnarray}
\label{tr}
&&-\frac{1}{c^2}\left[\frac{2}{a^2}\nabla^2 (V_N-U_N)\right]+\frac{1}{c^4} \Bigg\{-\frac{4}{a^2}\nabla^2(V_P-U_P)-\frac{2}{a^2}U_{N,k}U_N^{\phantom{N},k}-\frac{1}{a^2}V_{N,k} V_N^{\phantom{N},k}+\frac{2}{a^2} U_{N,k}V_N^{\phantom{N},k} \nonumber\\
&&\left.+\frac{4}{a^2}V_N\nabla^2(V_N-U_N)+6\left[\frac{\dot a}{a}(\dot U_N+3\dot V_N)+2\frac{\ddot a}{a}U_N+\left(\frac{\dot a }{a}\right)^2U_N+\ddot V_N\right]\right\} 
\nonumber\\&&=-\frac{8\pi G}{c^4}\bar\rho(1+\delta) v^2\;,
\end{eqnarray}
while the trace-free part is
\begin{eqnarray}
\label{tfb}
&&\frac{1}{c^2}\left[\frac{1}{a^2}(V_N-U_N)_{\;\;,i}^{,j}-\frac{1}{3a^2}\nabla^2 (V_N-U_N)\delta_{\;\;i}^{j}\right]+\frac{1}{c^4} \Bigg\{-\frac{1}{a}\frac{\dot a}{a}\left(B_{i}^{N,j}+B^{Nj}_{\phantom{Nj},i}\right)
-\frac{1}{2a}\left(\dot B_{i}^{N,j}+\dot B^{Nj}_{\phantom{Nj},i}\right)\nonumber\\&&+\frac{1}{a^2}\left[2(V_P-U_P)_{\;\;,i}^{,j}+U_{N,i}U_N^{\phantom{N},j}-V_{N,i}V_N^{\phantom{N},j}+U_{N,i}V_N^{\phantom{N},j}+U_N^{\phantom{N},j}V_{N,i} - 2V_N(V_N-U_N)_{\;\;,i}^{,j}+\frac{1}{2}\nabla^2h_{\;\;i}^{j}\right]\nonumber\\&&
+\frac{1}{a^2}\delta_{\;\;i}^{j}
\left[-\frac{2}{3}\nabla^2( V_P-U_P)-\frac{1}{3}(U_{N,k}U_N^{\phantom{N},k}- V_{N,k} V_N^{\phantom{N},k})-\frac{2}{3} U_{N,k} V_N^{\phantom{N},k}+\frac{2}{3}V_N\nabla^2(V_N-U_N)\right]\Bigg\}\nonumber\\&&
=-\frac{8\pi G}{c^4}\bar\rho(1+\delta)\left(v_iv^j -\frac{1}{3}\delta_{\;\;i}^{j}v^2\right).
\end{eqnarray}

\subsection{Scalar, vector and tensors parts}\label{SVTdecomp}
\subsubsection{Scalar equations}
It is useful to recast the previous equations in order to isolate, in the linear part of the equations, the scalar, vector and tensor contributions. For the scalar sector let us apply the divergence operator on Eq.\ (\ref{gi0}):
\begin{eqnarray}
\label{scalargi0}
&&\frac{1}{c^3}\nabla^2\left(\frac{\dot {a}}{a}U_N+\dot V_N\right)+\frac{1}{c^5}\bigg[2\nabla^2\left(\frac{\dot a}{a} U_P+\dot V_P\right)+\left(\dot V_N U_{,i}\right)^{,i}+2\frac{\dot a}{a} \left(U_NU_{,i}\right)^{,i}+2\left(\dot V_{N,i}V_N\right)^{,i}\nonumber\\
&&+\frac{1}{2a}B^N_{i\phantom{N},k}(V_N-U_N)^{,ki}
+\frac{1}{4a}\nabla^2B^N_i(V_N-3U_N)^{,i}+\frac{3}{4a}B^N_{i}\nabla^2 V_N^{\phantom{N},i}
\bigg]\nonumber\\
&&=\frac{4\pi G }{c^3}a \bar\rho \left[v_i(1+\delta)\right]^{,i}+\frac{4\pi G}{c^5}a\bar\rho \left\{\left[(\delta+1)v_i\left(v^2+2U_N+2V_N\right)\right]^{,i}-\delta^{,i} B^{N}_i\right\}\;.
\end{eqnarray}
Multiplying Eq.\ (\ref{scalargi0}) by $H/c$, and applying $\nabla^2$ on Eq.\ (\ref{ps}), we obtain the following constraint equation:
\begin{eqnarray}
\label{constraint}
&&\frac{1}{c^2}\nabla^2\nabla^2 V_N-\frac{1}{c^4}\left[\nabla^2\nabla^2 (V_N^2)-\frac{5}{2}\nabla^2(V_{N,i}V_N^{\phantom{N},i})-2\nabla^2\nabla^2V_P\right]
= -4\pi G (a^3 \bar\rho) \bigg\{\frac{1}{c^2} \frac{1}{a}\nabla^2\delta\nonumber\\
&&+\frac{1}{c^4}\left[ \frac{1}{a}\nabla^2((1+\delta) v^2)-3  \frac{\dot a}{a} (v_i(1+\delta))^{,i} \right]\bigg\}\;.
\end{eqnarray}
Now, we can obtain a second scalar constraint by applying the operator $\partial_j\partial^i$ on both sides of Eq.\ (\ref{tfb}). This then gives:
\begin{eqnarray}
\label{scalargij}
&&\frac{1}{c^2}\frac{2}{3}\nabla^2\nabla^2(V_N-U_N) +\frac{1}{c^4}\left\{\frac{4}{3}\nabla^2\nabla^2(V_P-U_P) + \left(U_{N,i}U_N^{\phantom{N},j}\right)_{,j}^{\;\;,i}-\left(V_{N,i}V_N^{\phantom{N},j}\right)_{,j}^{\;\;,i}+2\left(U_{N,i}V_N^{\phantom{N},j}\right)_{,j}^{\;\;,i}\right.\nonumber\\
&&\left.-\frac{1}{3}\nabla^2\left(U_{N,k}U_N^{\phantom{N},k}\right)+ \frac{1}{3}\nabla^2\left(V_{N,k} V_N^{\phantom{N},k}\right)-\frac{2}{3}\nabla^2\left(U_{N,k} V_N^{\phantom{N},k}\right)-2\left[V_N(V_N-U_N)_{,i}^{\;\;,j}\right]_{,j}^{\;\;,i}+\frac{2}{3}\nabla^2\left[ V_N\nabla^2(V_N-U_N)\right]\right\}\nonumber\\
&&=-\frac{8\pi G}{c^4}a^2\bar\rho\left[(1+\delta)\left(v_iv^j -\frac{1}{3}\delta_{i}^{j}v^2\right)\right]_{,j}^{\;\;,i}\;.\nonumber\\
\end{eqnarray}
These equations will be useful in Sec. \ref{NonlinearPN}.
\subsubsection{Vector equations}
The vectorial part of equation (\ref{gi0}) can be found using the curl operator:
\begin{eqnarray}
\label{vect}
&&-\frac{1}{c^3}\nabla\times\nabla^2B^N_i+\frac{1}{c^5}\nabla\times\left[-\nabla^2B^P_i+B^N_{i\phantom{N},k}(V_N-U_N)^{,k}-B^N_{k\phantom{N},i}(U_N+V_N)^{,k}+2\nabla^2B^N_{i}(V_N-U_N)+B^N_{i}\nabla^2 V_N\right.\nonumber\\&&\left.+2B^{N}_k V_{N,i}^{\phantom{N},k}\right]=\frac{16\pi Ga^2}{c^3}\nabla\times(\rho  v_i)+\frac{16\pi Ga^2}{c^5}\nabla\times\left(\rho \left\{v_i\left[v^2+2(U_N+V_N)\right]-B^N_i\right\}\right).
\end{eqnarray}
Alternatively, we can also obtain a  constraint equation for the vector part by applying  the operators $\nabla^2$  to  Eq.\ (\ref{gi0}) and $\partial_j$ to  Eq.\ (\ref{scalargi0}), finding
\begin{eqnarray}
\label{vect1}
&&\frac{1}{c^3} \nabla^2 \nabla^2B^N_{i}+\frac{1}{c^5}\bigg\{ \nabla^2 \nabla^2B^P_{i}-\nabla^2 \left[B^N_{i\phantom{N},k} \left(V_N-U_N\right)^{,k}\right]+\nabla^2 \left[B^N_{k\phantom{N},i} \left(V_N+U_N\right)^{,k}\right] -2 \nabla^2 \left[ \nabla^2 B^N_{i} \left(V_N-U_N\right)\right] \nonumber\\
&& - \nabla^2 \left(B^N_{i}\nabla^2V_N\right) -2 \nabla^2\left(B^N_{k}V_{N,i}^{\phantom{N},k}\right)+2\left[B^N_{j\phantom{N},k} \left(V_N-U_N\right)^{,jk}\right]_{,i}
+\left[\nabla^2 B^N_{j\phantom{N},k} \left(V_N-3U_N\right)^{,j}\right]_{,i}\nonumber\\
&&+3 \left(B^N_{j}\nabla^2V_N^{\phantom{N},j}\right)_{,i}+16\pi Ga^2 \bar\rho \left(\delta^{,j}B^N_{j}\right)_{,i}-16\pi G a^2 \bar\rho \;\nabla^2\left[\left(1+\delta \right) B^N_{j} \right] \bigg\}
=\frac{1}{c^3}16\pi G a^2  \bar\rho \left\{\left[v_j (1+\delta)\right]^{,j}_{,i}-\nabla^2\left[v_i (1+\delta)\right]\right\}\nonumber\\
&&+\frac{1}{c^5}\bigg\{4a\bigg[ \nabla^2\left(\dot V_N U_{N,i}\right)-\left(\dot V_N U_{N,k}\right)^{,k}_{,i}+2\frac{\dot a }{a}\nabla^2\left(U_N U_{N,i}\right)-2\frac{\dot a }{a}\left(U_N U_{N,k}\right)^{,k}_{,i} +2  \nabla^2\left(\dot V_{N,i} V_{N}\right)-2\left(\dot V_{N,k} V_{N}\right)^{,k}_{,i}\bigg] \nonumber\\
&&+16\pi G a^2 \bar\rho \bigg[v_j (1+\delta)\left(v^2+2U_N+2V_N\right)^{,j} \bigg]_{,i}-16\pi G a^2 \bar\rho\; \nabla^2 \bigg[ v_i (1+\delta)\left(v^2+2U_N+2V_N\right)\bigg]\bigg\}\;.
\end{eqnarray}

These vectorial equations  also depend on scalar  quantities, because of nonlinearity; however, the divergence of these equations would give zero, as it should. 

An evolution equation for $B_i$ can be obtained by applying the operator $\partial_j\nabla^2$ on  Eq.\ (\ref{tfb}) and subtracting the equation obtained applying $\partial_i$ on  Eq.\ (\ref{scalargij}). This procedure leads to 
\begin{eqnarray}
\label{dynvect}
&&\frac{1}{c^4}\nabla^2\nabla^2\left(\frac{1}{2}\dot B^N_{i} +\frac{\dot a}{a} B^N_{i}\right) = \frac{1}{c^4}\frac{1}{a}
\bigg\{\nabla^2\left(U_{N,i} U_{N}^{\phantom{N},j}\right)_{,j}-\left(U_{N,k}U_{N}^{\phantom{N},j}\right)_{,ji}^{\;\;\;,k}-\nabla^2\left(V_{N,i}V_{N}^{\phantom{N},j}\right)_{,j}+\left(V_{N,k}V_{N}^{\phantom{N},j}\right)_{,ji}^{\;\;\;,k}\nonumber\\
&&+\nabla^2\left(U_{N,i}V_{N}^{\phantom{N},j}\right)_{,j}+\nabla^2\left(U_N^{\phantom{N},j}V_{N,i}\right)_{,j}-2\left(U_{N,k}V_{N}^{\phantom{N},j}\right)_{,ji}^{\;\;\;,k}-2\nabla^2\left[(V_N-U_N)_{,i}^{,j}V_{N}\right]_{,j}+2\left[(V_N-U_N)_{,k}^{,j}V_{N}\right]_{,ji}^{\;\;\;,k}\bigg\}\nonumber\\
&&+\frac{8\pi G}{c^4}a\bar\rho\left\{\nabla^2\left[(1+\delta)(v_iv^k -\frac{1}{3}\delta_{i}^{k}v^2)\right]_{,k}-\left[(1+\delta)(v_kv^j -\frac{1}{3}\delta_{k}^{j}v^2)\right]_{,ji}^{\;\;\;,k}\right\}\;.
\end{eqnarray}
This shows that an evolution term for $B^N_i$ only appears at order $1/c^4$.
\subsubsection{Tensor equations}\label{sec-TT}
In order to isolate the TT part of the metric $h_{ij}$  we define the following nonlinear quantities:
 \begin{eqnarray}\label{A}
\mathcal{A}_i^{\;\;j}&=&U_{N,i}U_N^{\phantom{N},j}-V_{N,i}V_N^{\phantom{N},j}+U_{N,i}V_N^{\phantom{N},j}+U_N^{\phantom{N},j}V_{N,i}-2V_N(V_N-U_N)_{,i}^{\;\;,j}\nonumber\\
&+&\delta_i^j\bigg[-\frac{1}{3}U_{N,k}U_{N}^{\phantom{N},k}+\frac{1}{3} V_{N,k} V_{N}^{\phantom{N},k}-\frac{2}{3}U_{N,k} V_{N}^{\phantom{N},k}+\frac{2}{3}V_N\nabla^2(V_N-U_N)\bigg]\;, \\
&& \nonumber
 \\ \label{B}
\mathcal{S}_i^{\;\;j}&=&(1+\delta)\left(v_iv^j -\frac{1}{3}\delta_{i}^{j}v^2\right)\;.
\end{eqnarray}
With these definitions, Eq.(\ref{dynvect}) becomes
\begin{eqnarray}
\label{dynvects2}
&&\frac{1}{c^4}\nabla^2\nabla^2\left(\frac{1}{2a}\dot B^N_{i} +\frac{1}{a}\frac{\dot a}{a} B^N_{i}\right) = \frac{1}{c^4}\left[\frac{1}{a^2}\left(\nabla^2\mathcal{A}_{i,j}^{j}-\mathcal{A}_{k,ji}^{i,k}\right)+8\pi G \bar\rho \left(\nabla^2\mathcal{S}_{i,j}^{j}-\mathcal{S}_{k,ji}^{i,k}\right)\right]\text{.}
\end{eqnarray}
Finally, using Eqs. (\ref{tfb}), (\ref{scalargij}) and  (\ref{dynvects2}), we  obtain the following  constraint equation for $h_{ij}$:
\begin{eqnarray}
\label{hij}
&&\frac{1}{c^4}\nabla^2\nabla^2\nabla^2h_{i}^{j}=\frac{1}{c^4}\left[-\mathcal{A}_{k,li}^{l,kj}-\nabla^2\mathcal{A}_{k,l}^{l,k}\delta_{i}^{j}+2\nabla^2\mathcal{A}_{i,k}^{k,j}   +2\nabla^2\mathcal{A}^{k}_{l,ki} \delta^{lj}-2\nabla^2\nabla^2\mathcal{A}_i^j\right.\nonumber\\
&&\left.+8\pi G a^2 \bar\rho\left(-\mathcal{S}_{k,li}^{l,kj}-\nabla^2\mathcal{S}_{k,l}^{l,k}\delta_i^j+2\nabla^2\mathcal{S}_{l,ki}^{k} \delta^{lj}+2\nabla^2\mathcal{S}^{k,j}_{i,k}-2\nabla^2\nabla^2\mathcal{S}_i^j\right)\right]\;.
\end{eqnarray}

\section{Conservation equations}\label{ConservationEq}

The field  equations in GR are constructed in order to imply the conservation equations through the contracted Bianchi identities  \cite{Einstein}:
\begin{equation}
\label{conss}
T_{\;\;\mu;\nu}^{\nu}=0\;.
\end{equation}

From this, considering the time component and keeping all terms up to $1/c^2$ order, we obtain the energy conservation equation \\
\begin{eqnarray}
\label{contin}
\frac{(a^3\rho)^\cdot{}}{a^3}+\frac{(v^i\rho)_{,i}}{a}+\frac{1}{c^2}\left[\frac{(a^4\rho v^2)^\cdot{}}{a^4}+3 \dot V_N \rho +\frac{(\rho v^2 v^i)_{,i}}{a}+\rho v^i\frac{(3V_N-U_N)_{,i}}{a}\right]=0\;.
\end{eqnarray}
Setting $v^i=U_N=V_N=0$, and $\rho=\bar{\rho}$,  this equation reduces to the background continuity equation for  cold dark matter, i.e. $\dot{\bar\rho}=-3H\bar\rho$.
Subtracting this equation from (\ref{contin}) and defining the total, or convective, derivative for any quantity $Q$ as 
\begin{equation}
\frac{dQ}{dt}=\dot Q +\frac{v^iQ_{,i}}{a}\text{,}
\end{equation}
 Eq.~(\ref{contin}) becomes\\
\begin{eqnarray}
\label{continu}
\frac{d\delta}{dt} +\frac{v^i_{\;,i}}{a}(\delta+1)+\frac{1}{c^2}\left\{(\delta+1)\left[\frac{\dot a}{a}v^2+\frac{d(v^2)}{dt}+3\frac{dV_N}{dt}-\frac{v^i}{a}U_{N,i}\right]\right\}=0\;.
\end{eqnarray}
Note that  we have used the  Newtonian part of this equation to simplify the 1PF order part. 
The space part of (\ref{conss}) gives the momentum conservation equation:\\
\begin{eqnarray}\label{euler1PF}
\frac{(a^4\rho v_i)^\cdot{}}{a^4}-\frac{\rho U_{N,i}}{a}+\frac{(v^j\rho v_i)_{,j}}{a}+\frac{1}{c^2}\left\{\rho v_iv^j\frac{(3V_N-U_N)_{,j}}{a}-2\frac{\rho U_{P,i}}{a}+\rho v_i(3V_N-U_N)^\cdot{}+\frac{(\rho v_iv^jv^2)_{,j}}{a}+\frac{(a^4\rho v^2v_i)^\cdot}{a^4}+\right.\nonumber\\
+\left.\frac{\left[2a^4 \rho v_i(V_N+U_N)\right]^\cdot}{a^4}+2\frac{\left[\rho v^j v_i(V_N+U_N)\right]_{,j}}{a}-\rho v^2 \frac{(V_N+U_N)_{,i}}{a}-\frac{(a^4 B^N _i\rho)^{\cdot}}{a^4}-\frac{(B^N_i\rho v^j)_{,j}}{a}+\frac{B^N_{j,i}}{a}\rho v^j\right\}=0\;.\nonumber\\
\end{eqnarray}
Note that Eqs.\ (\ref{contin}) and (\ref{euler1PF}) are equivalent to Eqs.\ (57) and (58) in \cite{Hwang:2005mg}: as far as the conservation equations are concerned, there is no difference between the standard post-Newtonian approach used in \cite{Hwang:2005mg} and our post-Friedmann approach. The difference becomes relevant for the consistency of the full set of Einstein equations; we discuss this point in Sec. \ref{pas-act}.
Simplifying  this equation by using the   Newtonian part of Eq.~(\ref{continu}) and the background continuity equation for $\bar\rho$,  we derive the 1PF Euler equation:\\
\begin{eqnarray}
\label{eulero}
&&\frac{dv_i}{dt}+\frac{\dot a}{a}v_i-\frac{U_{N,i}}{a}+\frac{1}{c^2}\bigg[v_i\frac{d}{dt}(U_N+2V_N)+\frac{2}{a}U_{N,i}(U_N+V_N)-\frac{1}{a}v^2V_{N,i}-\frac{2}{a}U_{P,i}+\frac{1}{a}v_iv^jU_{N,j}-\frac{\dot a}{a}v^2v_i\nonumber\\
&&-\frac{1}{a}\frac{d}{dt}(a B^N _i)+\frac{B^N_{j,i}v^j }{a}\bigg]=0\;.
\end{eqnarray}
Finally, using the   Newtonian part of Eq. (\ref{eulero}), the continuity equation can be recast in the following way:
\begin{eqnarray}
\label{continuity}
\frac{d\delta}{dt} +\frac{v^{i}_{\;,i}}{a}(\delta+1)+\frac{1}{c^2}\left[(\delta+1)\left(\frac{1}{a}v^jU_{N,j}-\frac{\dot a}{a}v^2+3\frac{dV_N}{dt}\right)\right]=0\;.
\end{eqnarray}

The last two equations  are the hydrodynamic equations of motion of the dust component  in the post-Friedmann approximation, at 1PF order. 
Let us stress that, in Eqs.\ (\ref{continuity}) and (\ref{eulero}), the post-Friedmann corrections contain nonlinear coupling terms between metric variables and matter variables, $\rho$ and $v_i$. This is the reason why the vector potential $B_i$ cannot be decoupled from the scalar modes in the equations of motion. 

In the next two sections we will check the consistency of our approach with our goal of obtaining a set of equation valid at all scales. To this end we will first  consider the leading order, the Newtonian regime, which is a good approximation on scales much smaller than the Hubble radius, then we will linearize our equation to check consistency with   linear relativistic perturbation theory in the Poisson gauge \cite{Ma:1995ey,Bertschinger:1993xt},  which is a good approximation on large scales.

\section{Leading order: the Newtonian regime}\label{Newtonianregime}
\subsection{Newtonian dynamics from consistency of Einstein equations, with a bonus}
 
 Retaining the leading order-terms in the $c^{-1}$ expansion, i.e.\ the 0PF order, we  recover the equations of Newtonian cosmology and we obtain the corresponding spacetime metric. We call this the Newtonian regime: as we are going to see, the dynamics is purely Newtonian, yet we have a spacetime metric that is a well-defined approximate solution of Einstein equations. 
 
We obtain the Newtonian continuity and Euler equation from the hydrodynamic equations (\ref{continu}) and (\ref{eulero}):

\begin{eqnarray}
\label{d}
&&\dot\delta +\frac{v^i\delta_{,i}}{a}+\frac{v^i_{\;\;,i}}{a}(\delta+1)=0 \;,\\
\label{e}
&&\dot v_i +\frac{v^jv_{i,j}}{a}+\frac{\dot a}{a} v_i=\frac{1}{a}U_{N,i}\;.
\end{eqnarray}\\ 
At leading order, from Einstein equations (\ref{ps}), (\ref{gi0}), (\ref{tr}), (\ref{tfb}), we obtain
\begin{subequations}
\begin{eqnarray}
\label{a}
G_{\;\;0}^0+\Lambda=\frac{8\pi G}{c^4}T^0_{\;\;0}&\rightarrow&\;\frac{1}{c^2}\frac{1}{a^2}\nabla^2 V_N=-\frac{4\pi G}{c^2}\bar\rho\delta\;,\\
\label{aa}
G^0_{\;\;i}=\frac{8\pi G}{c^4}T^0_{\;\;i}&\rightarrow&\;\frac{1}{c^3}\left[-\frac{1}{2a^2}\nabla^2B^N_{i}+2\frac{\dot a}{a^2}U_{N,i}+\frac{2}{a}\dot V_{N,i}\right]=\frac{8\pi G}{c^3}\bar\rho(1+\delta) v_i\;,\\
\label{b}
{\rm trace\; of\;\; } G^i_{\;\;j}+\Lambda \delta^i_{\;\;j}=\frac{8\pi G}{c^4}T^i_{\;\;j}&\rightarrow&\;\frac{1}{c^2}\frac{2}{a^2}\nabla^2 (V_N-U_N)=0\;,\\
\label{ci}
{\rm traceless\; part\; of}\;\; G^i_{\;\;j}+\Lambda\delta^i_{\;\;j}=\frac{8\pi G}{c^4}T^i_{\;\;j}&\rightarrow&\;\frac{1}{c^2}\frac{1}{a^2}\bigl[(V_N-U_N)_{,i}^{\;\;,j}-\frac{1}{3}\nabla^2 (V_N-U_N)\delta_{i}^{j}\bigr]=0\;.
\end{eqnarray}\\
\end{subequations}
Equation (\ref{a}) is the Poisson equation, from which the Newtonian character of the spatial metric potential $V_N$ is apparent as long as we identify it with the Newtonian gravitational potential generated by the matter field, $V_N=-\phi_N$. As noted by Bertschinger \cite{mit-notes}, it is an interesting fact that at leading order only $g_{ij}$ contributes to $G^{0}_{\;\;0}$. On the other hand, as is well known, it is the time-time metric potential $U_N$ that tells matter how to move in the Euler equation (\ref{e}) (or to particles in the geodesic equation; see e.g.\ \cite{Misner:1973mi}). Equations (\ref{b}) and (\ref{ci}) show that Eqs.\ (\ref{tr}) (\ref{tfb}) reduce, at leading order, to constraint equations: they give the consistency relations between the scalar metric potentials $U_N$ and $V_N$ in a GR context at this order. Modulo residual gauge modes \cite{mit-notes} that leave (\ref{b}) and (\ref{ci}) invariant, we must have $U_N=V_N=-\phi_N$. In summary,  Einstein equations reduce, at leading order,  to the standard equations of Newtonian cosmology.
The metric tensor generated  from a self-consistent expansion of the full set of Einstein equations at leading order is the  cosmological version of the weak field metric, with the FLRW metric replacing Minkowski as background. 

Our metric (\ref{metric}), however, also contains the frame dragging vector potential $B_i$ and the TT part $h_{ij}$. This TT part only appears at 1PF order in $G^i_{j}$ and so it is irrelevant at leading order, consistently with the fact that, in this Newtonian regime, we only retain the $V_N$ term in $g_{ij}$. 
On the other hand, the leading order of the $G^0_i$ equation is Eq.\ (\ref{aa}); thus, it can not be neglected. This is a new equation that arises in the relativistic context and
determines the leading-order frame dragging potential $B^N_i$ from the purely Newtonian term on the right hand side, the energy current that is determined by the other equations. Therefore, even in the Newtonian regime, the  frame dragging term $B^N_i$ cannot be set to zero. Some authors consider this term of higher order; however Eq.\ (\ref{aa}) tell us that this is inconsistent with the other Newtonian equations. Indeed, taking the divergence of (\ref{aa}) (i.e. its scalar part) and using the Poisson equation implies the continuity equation, while taking the curl (i.e.\ the vector part) shows that the curl of $B^N_i$,  a gravitomagnetic field \cite{mit-notes}, is sourced by the curl of the energy current $\rho v_i$, and there is no reason why the transverse part of this current should vanish. Alternatively \cite{mit-notes}, one can split vectors into  longitudinal and  transverse parts. 
Then one can see from (\ref{aa}) that a purely Newtonian transverse energy current sources $B^N_i$. Thus, setting $B^N_i=0$  does imply, even  in the Newtonian regime of dynamics, a purely longitudinal energy current, i.e.\  imposing the condition $B^N_i=0$ does imply an artificial constraint on the general Newtonian dynamics. While in the linear regime  vectors modes can be separated from  scalar modes, in the nonlinear regime of Newtonian dynamics there is no reason why the energy current should be longitudinal, and indeed it is well known that transverse vector modes are generated by nonlinearity, and in particular vorticity is generated
after shell crossing \cite{Pueblas:2008uv}.
The nonlinearly generated gravitomagnetic potential $B^N_i$ can actually be extracted from standard Newtonian N-body simulations \cite{Bruni:2013mua,2015arXiv150100799T}, and its contribution to lensing has been computed in \cite{Thomas:2014aga}. The bottom line is that in the Newtonian regime of GR the gravitational field is not purely scalar, it does contain a gravitomagnetic vector part \cite{Ehlers:2009uv,Kofman:1994kf}.

As we are going to show in the next section, the above considerations on the frame dragging $B_i$ are entirely consistent with linear perturbation theory. In the perturbative regime one naturally splits vectors into longitudinal and transverse parts and finds that $B_i$ decays \cite{Bardeen:1980kt,Bruni:1992dg,Szekeres:2000ki,Szekeres:1999ny,Carbone:2004iv}.
Nonetheless, in the nonlinear Newtonian regime the transverse part of the energy current sources $B^N_i$ \cite{Bruni:2013mua,2015arXiv150100799T}. At this order, however, $B^N_i$ is a nondynamical variable, rather it is determined nonlocally  through the acausal equation (\ref{aa}). Furthermore, a calculation of the magnetic Weyl tensor shows that, at leading order in powers of $c^{-1}$,
\begin{equation}\label{magweyl}
H_{ij}=\frac{1}{2c^3}\left[ B^N_{\mu,\nu(i}\varepsilon_{j)}^{\;\;\;\;\mu\nu}+2v_\mu(U_N+V_N)_{,\nu(i}\varepsilon_{j)}^{\;\;\;\;\mu\nu}\right],
\end{equation} 
so that its linear part, i.e.\ the first term in the square bracket, is precisely the curl of $B^N_i$.
As discussed in detail by Kofman and Pogosyan \cite{Kofman:1994kf}, although the magnetic Weyl tensor is zero in Newtonian theory (i.e.\ starting from a scalar theory of gravity), it cannot be neglected in deriving the ``covariant equations'' \cite{ellis2012relativistic,ellis1971Varenna} in the Newtonian regime from relativistic theory.
Nonetheless $H_{ij}$ is not an independent dynamical variable in the Newtonian regime, rather Eq.\ (\ref{magweyl}) shows that it is locally determined by other variables.

\subsection{Passive and active approach}\label{pas-act}

Starting with Einstein \cite{Einstein}, the usual way to consider the Newtonian limit in GR consists in demanding that the timelike geodesic equation agrees with the Newtonian equation of motion for a particle \cite{Weinberg:1972zz,Misner:1973mi}. This naturally sets $g_{00}=-(1-2U_N/c^2)$, identifying $-U_N$ with the gravitational potential $\phi_N$. For a fluid, as we can see from (\ref{d}) and (\ref{e}), the continuity and the Euler equations do not contain $V_N$, so that the latter could be interpreted as a post-Newtonian variable. This way of thinking considers the ``passive'' aspect of gravitation, namely the response of matter to gravity \cite{Misner:1973mi}: one wants to determine the equation of motion of a particle in a \itshape given \normalfont  gravitational potential. 

In addition, the Newtonian limit is again considered using the field equations with matter, in order to determine the gravitational coupling constant by recovering the Poisson equation \cite{Einstein}. Starting from the following form of  Einstein equations
\begin{equation}
\label{altra}
R_{\mu\nu}=\kappa(T_{\mu\nu}-\frac{1}{2}T\delta _{\mu\nu})+\Lambda \delta_{\mu\nu}\,
\end{equation}
one finds that the time-time component only contains $g_{00}$, so that, at $O(c^{-2})$, we obtain the Poisson equation for $U_N$\\
\begin{equation}
\label{ff}
\frac{1}{c^2}\frac{1}{3a^2}\nabla^2 U_N=-\frac{4\pi G}{3c^2}\bar\rho\delta \,
\end{equation}\\
by identifying $\kappa =8\pi G /c^4$. Again, one would be tempted to conclude that only $U_N$ is needed at leading Newtonian order.

However, in our approach, we are interested in the ``active'' aspect of gravitation, i.e. the generation of gravity by the matter distribution. To this end, we need to determine the metric using the field equation in a fully consistent manner, even in the Newtonian regime. 
In particular, in cosmology we deal with  a self-gravitating fluid, so that the complete set of field equations is as important as the equations of motion. 

In the passive Newtonian approach of the standard post-Newtonian formalism the geodetic equation determines the order of the metric variables, $U_N$ as Newtonian and $V_N$ as the post-Newtonian correction. In our post-Friedmann scheme it is the leading order of the  field equations, derived from  the complete set  of  Einstein equation, that establishes the  order of the metric perturbations,  giving that both $U_N$ and $V_N$ have a Newtonian character.
 This emerges from (\ref{a}), (\ref{b}) and (\ref{ci}) and naturally also from the spatial components of (\ref{altra}), whose trace is
\begin{eqnarray}\label{g}
\frac{1}{c^2}\frac{1}{a^2}\nabla^2(U_N-4V_N)=\frac{12\pi G}{c^2}\bar\rho\delta\;.
\end{eqnarray}\\
This clearly shows that neglecting $V_N$ with respect to $U_N$ would not yield the Poisson equation. 

In summary, if we want to obtain the equations of Newtonian cosmology from GR, in the active approach and consistently considering all  components of  Einstein equations, we do need  both  scalar potentials $U_N$ and $V_N$ as 0PF terms in the metric, with the result 
\begin{equation}\label{Npot}
U_N=V_N=-\phi_N\;.
\end{equation}

\section{Linearization: recovering first-order perturbation theory}\label{linear_limit}

Standard relativistic perturbation theory \cite{Bardeen:1980kt,Kodama:1985bj,Bruni:1992dg,Bertschinger:1993xt,Malik:2008im,Uggla:2012gg}, where density, velocity and metric variables are all assumed to be small, is a fundamental tool in theoretical cosmology. It provides the framework to develop predictions from inflation for the matter and metric fluctuations in the early Universe and to work out their imprint  on the CMB 
and the matter fluctuations at the beginning of the matter dominated era \cite{Bruni:2013qta}. 

Our post-Friedmann scheme has been developed to generalize the equations of nonlinear Newtonian cosmology to include relativistic corrections. It would, however, be a great bonus if, in linearizing our equations, we could recover standard first-order relativistic perturbation theory. Given that the observed Universe is remarkably well described by perturbation theory at large scales, up to the Hubble horizon and larger, recovering at least the first order would imply that our scheme is not only valid at small nonlinear Newtonian scales, as shown in the previous sections, but also at the largest scales of interests in cosmology.  Note that it is not {\it a priory} obvious that this recovery is possible; the aim of this section is to explicitly show that, by linearizing our equations and by defining resummed metric variables that contain a Newtonian part and a relativistic 1PF correction, we obtain first-order perturbation theory. This resummation will also be at the base of the following sections.   

In  first-order perturbation theory, it is standard to decouple the scalar, vector and tensor modes  \cite{Bardeen:1980kt,Kodama:1985bj,Bruni:1992dg,Bertschinger:1993xt,Ma:1995ey,Malik:2008im}. It is, therefore, convenient to separately  look at the scalar and vector type equations of Sec.\ \ref{SVTdecomp} (as already noted, at 1PF order the tensorial modes are not dynamical, i.e.\ we can't recover first-order gravitational waves).
In particular, the first-order velocity perturbation $v^i$ can also be split into a scalar and vector (solenoidal) part, $v^i=v^{\|\;,i} +v^i_{\perp}\;$, where $v^i_{\perp\;,i}=0$.
 
Starting from the scalar sector and considering only the linear terms, we obtain from equations (\ref{ps}), (\ref{tr}), (\ref{scalargi0}), and (\ref{scalargij})
\begin{subequations}
\begin{eqnarray}
\label{pslin}
&&\frac{1}{c^2}\frac{1}{3}\nabla^2 V_N-\frac{1}{c^4}a^2\left[\frac{\dot a}{a}\dot V_N+\left(\frac{\dot a}{a}\right)^2U_N-2\frac{\nabla^2V_P}{3a^2}\right]
=-\frac{1}{c^2}\frac{4\pi G}{3}a^2\bar\rho \delta \;,\\
\label{trlin}
&&\frac{1}{c^2}\left[\nabla^2 (U_N-V_N)\right]+\frac{1}{c^4}\left\{2\nabla^2\left(U_P-V_P\right)+ 3a^2\left[\frac{\dot a}{a}\left(\dot U_N+3\dot V_N\right)+2\frac{\ddot a}{a}U_N+\left(\frac{\dot a }{a}\right)^2U_N+\ddot V_N\right]\right\}=0\;, \nonumber \\ \label{trlin1} \\
\label{scalargi0lin}
&&\frac{1}{c^3}\nabla^2\left(\frac{\dot {a}}{a}U_N+\dot V_N\right)+\frac{1}{c^5}\left(
2\frac{\dot a}{a} \nabla^2U_P+2\nabla^2\dot V_P\right)=\frac{1}{c^3}4\pi G a \bar\rho \theta\;,\\
\label{scalargijlin}
&&\frac{1}{c^2}\nabla^2\nabla^2(V_N-U_N)+ \frac{1}{c^4}2\nabla^2\nabla^2(V_P-U_P)=0\;,
\end{eqnarray}
\end{subequations}
where, in Eq.\ (\ref{scalargi0lin}), we have defined the spatial velocity divergence as $\theta=v^{i}_{\;,i}=\nabla^2v^{\|}\;$.
The continuity equation and the divergence of the Euler equation from (\ref{continuity}) and (\ref{eulero}), after linearization, are
\begin{eqnarray}
\label{continulin}
&&\dot\delta  +\frac{\theta}{a}+\frac{3}{c^2}\dot V_N=0\;,\\
\label{eulerolin}
&&\dot \theta+\frac{\dot a}{a}\theta-\frac{\nabla^2U_N}{a}-\frac{2}{c^2}\frac{\nabla^2U_P}{a}=0 \;,
\end{eqnarray}
where the convective term in the total time derivatives is neglected and partial and total derivatives coincide.
Finally, defining the following resummed scalar metric variables 
\begin{eqnarray}
\label{phi}
&&\phi_P:=-(U_N+\frac{2}{c^2}U_P),\\
\label{psi}
&&\psi_P:=-(V_N+\frac{2}{c^2}V_P),
\end{eqnarray}
the previous equations become
\begin{subequations}
\begin{eqnarray}
\label{pslinp}
&&\nabla^2 \psi_P-\frac{3}{c^2}a^2\left[\frac{\dot a}{a}\dot \psi_P+\left(\frac{\dot a}{a}\right)^2\phi_P\right]=4\pi G\bar\rho a^2 \delta\;, \\
\label{trlinnew}
&&-\nabla^2(\psi_P-\phi_P)+ \frac{3}{c^2} a^2 \left[\frac{\dot a}{a}(\dot \phi_P+3\dot \psi_P)+2\frac{\ddot a}{a}\phi_P+\left(\frac{\dot a }{a}\right)^2\phi_P+\ddot \psi_P\right]=0\;,\\
\label{scalargi0lin2}
&&\nabla^2\left(\frac{\dot {a}}{a}\phi_P+\dot \psi_P\right)=-4\pi G a \bar\rho \theta\;,\\
\label{hu}
&&\frac{1}{c^2}\nabla^2\nabla^2(\phi_P-\psi_P)=0\;,\\
\label{continulinnew}
&&\dot\delta  +\frac{\theta}{a}-\frac{3}{c^2}\dot \psi_P=0\;,\\
\label{eulerolinnew}
&&\dot \theta+\frac{\dot a}{a}\theta+\frac{1}{a}\nabla^2\phi_P=0\;.
\end{eqnarray}
\end{subequations}
Note that, from Eq.\ (\ref{hu}),  $\psi_P=\phi_P$; this follows from the perfect fluid matter model we are considering (i.e.\ null anisotropic stress, see e.g.\ \cite{Bardeen:1980kt,Bruni:1992dg}).
The equations above coincide with those for scalar fluctuations of standard perturbation theory in the Newtonian gauge; see e.g.\ \cite{Bertschinger:1993xt,Ma:1995ey,poisgaug}.
The conclusion of this analysis is important: the 1PF approximation, in the linear regime, completely includes  first-order scalar relativistic perturbation theory, so that all the linear GR terms are just 1PF. On the other hand, the 1PF order  does not involve just linear terms, indeed there are contributions of second and higher order in the standard perturbative expansion.

This result is due to our active approach, where $U_N$ and $V_N$ are both Newtonian variables,  to the  derivation of 1PF equations that retain the first two orders in the $1/c$ expansion (rather than proceeding  iteratively, order by order), and to the definition of the resummed variables (\ref{phi}) and (\ref{psi}) that turn out to coincide with the standard first-order potentials in  Poisson gauge.

We now briefly consider  the linear vector sector. Again defining a resummed variable, let us consider the  vectorial potential $\omega_i$:
\begin{equation}
\label{newvect}
\omega_i=B^N_i+\frac{1}{c^2}B^P_i.
\end{equation}
 Linearizing Eqs.\ (\ref{vect}) and (\ref{tfb}), we immediately obtain
\begin{eqnarray}
\label{linearvectorg0i}
\frac{1}{c^3}\nabla^2{\omega}^{i}=-\frac{1}{c^3}16\pi G a^2\bar\rho \left(v_{\perp}^{i}- \frac{1}{c^2} \omega^i\right)\;,\\
\label{linearvectorgij}
2\frac{\dot a}{a}\left(\omega_{i}^{\;,j}+\omega^{j}_{\;,i}\right)+\left(\dot \omega_{i}^{\;,j}+\dot{\omega}^{j}_{\;,i}\right)=0\;.
\end{eqnarray}
Therefore, the linearization of the 1PF approximation also reproduces the equations of  linear relativistic  perturbations theory in the vector sector. Note that Eq.\ (\ref{linearvectorg0i}) is the first-order analogue of the nonlinear Newtonian Eq.\ (\ref{aa}). Eq.\ (\ref{linearvectorgij}) instead gives the evolution of the first-order metric vector potential, which decay.
In the nonlinear case, it is indeed shell-crossing that generates  vorticity \cite{Pueblas:2008uv}. In general, nonlinearity generates the vector modes  in the energy current that sources the gravitomagnetic frame-dragging potential \cite{Bruni:2013mua,2015arXiv150100799T}.

Finally,  a remark on the tensor sector. Had we extended our post-Friedmann scheme to the 2PF order, in the linear regime we would have recovered the wave equation for free gravitational waves.  In the nonlinear regime gravitational radiation would be generated by nonlinear sources terms, of higher order in scalar and vector modes, much in the same way that is generated at second order in standard perturbation theory \cite{Bruni:1996im,poisgaug,Ananda:2006af,Malik:2008im}. 

\section{nonlinear set of 1PF equations} \label{NonlinearPN}
\subsection{Resummed metric potentials and new conservation equations}
 
 We now recast the nonlinear  equations obtained in the previous sections, using a suitable change of variables. In the previous section we have introduced the potentials $\phi_P $ and $\psi_P$:  we have shown that they coincide in both  the linear and  the Newtonian regime. It is then convenient to consider the following combinations:
\begin{eqnarray}
\phi_G&:=&\frac{1}{2}(\phi_P+\psi_P),\\
\frac{1}{c^2}D_P&:=&\frac{1}{2}(\phi_P-\psi_P);
\end{eqnarray}
the first generalizes the definition of the gravitational potential given in Sec.\ \ref{Newtonianregime}, equation (\ref{Npot}), the second defines a new nonlinear post-Friedmannian quantity, $D_P$, negligible in the Newtonian and in the linear regimes\footnote{A difference in the two scalar potentials is expected from nonlinearity in GR, consistently with second-order perturbation theory results \cite{poisgaug,Malik:2008im,Uggla:2012gg}.}. 
With these new variables, up to $O(1/c^4)$, the metric can be written as
\begin{eqnarray}
ds^2=-c^2\left[1+2 \left(\frac{\phi_G}{c^2}+\frac{\phi_G^2}{c^4}+\frac{D_P}{c^4}\right)\right]dt^2-2a\frac{\omega_i}{c^2}dtdx^i  + a^2\left\{\left[1+2\left(-\frac{\phi_G}{c^2}+\frac{\phi_G^2}{c^4}+\frac{D_P}{c^4}\right)\right]\delta_{ij}+\frac{1}{c^4}h_{ij}\right\}dx^i dx^j\;. \nonumber\\ 
\end{eqnarray}
Moreover, it is convenient to define a new velocity variable \cite{Chandrasekhar:1965ch},
\begin{eqnarray}
\label{v}
 v^*_i=v_i-\frac{1}{c^2}\omega_i\;,
\end{eqnarray}
representing the velocity of matter with respect to observers moving along the normal to the slicing.
With these new variables, Eqs.\ (\ref{ps}), (\ref{tr}), (\ref{tfb}) and (\ref{gi0}) become
\begin{subequations}
\label{lll}
\begin{eqnarray}
\label{psp}
G_{\;\;0}^0+\Lambda=\frac{8\pi G}{c^4}T^0_{\;\;0}\quad\rightarrow\;&&\frac{1}{c^2}\nabla^2\phi_G-\frac{1}{c^4}\left[\nabla^2D_{P}+3a^2\left(\frac{\dot a}{a}\dot \phi_G+\left(\frac{\dot a}{a}\right)^2\phi_G\right)-\nabla^2 \phi_G^2+\frac{5}{2}\phi_{G,i}\phi_{G}^{\;\;\;,i}\right]\nonumber\\
&&=\frac{1}{c^2}4\pi Ga^2\bar\rho \delta+\frac{1}{c^4}4\pi Ga^2\bar\rho(1+\delta)v^{*2},\\
\label{trp}
\left[G_{\;\;i}^j+\Lambda\delta_i^j=\frac{8\pi G}{c^4}T^j_{\;\;i}\right]_{\rm trace}\rightarrow\; &&
\frac{1}{c^4}\left[4\nabla^2 D_{P}+\phi_{G,k} \phi_{G}^{\;\;\;,k}+6a^2\left(4\frac{\dot a}{a}\dot \phi_G+2\frac{\ddot a}{a}\phi_G+\left(\frac{\dot a }{a}\right)^2\phi_G+\ddot \phi_G\right)\right]\nonumber\\&&=\frac{1}{c^4}8\pi Ga^2\bar\rho(1+\delta) v^{*2},\\
\label{tfbp}
\left[G_{\;\;i}^j+\Lambda\delta_i^j=\frac{8\pi G}{c^4}T^j_{\;\;i}\right]_{\rm tracefree}\rightarrow\; &&
\frac{1}{c^4}\left\{2\left(D_{P,i}^{\;\;\;,j}-\frac{1}{3}\nabla^2D_{P}\,\delta_i^j\right)+2\phi_{G,i}\phi_G^{\;\;\;,j}-\frac{2}{3}\phi_{G,k}\phi_{G}^{\;\;\;,k}\,\delta_i^j\right.\nonumber\\&&\left.-a\left[\frac{\dot a}{a}\left(\omega_{i}^{\;,j}+\omega^{j}_{\;\;,i}\right)+\frac{1}{2}\left(\dot \omega_i^{\;,j}+\dot \omega^{j}_{\;\;,i}\right)\right]+\frac{1}{2}\nabla^2h_{i}^{j}\right\}\nonumber\\&&=-\frac{1}{c^4}8\pi G a^2\bar\rho(1+\delta)\left(v^*_iv^{*j}-\frac{1}{3}v^{*2}\,\delta_i^j\right),\\\nonumber\\
\label{gi0p}
G_{\;\;i}^0=\frac{8\pi G}{c^4}T^0_{\;\;i}\quad\rightarrow\;&& 
\frac{1}{c^3}\left[\frac{1}{2a}\nabla^2\omega_i+2\frac{\dot a}{a} \phi_{G,i}+2\dot \phi_{G,i}\right]-\frac{1}{c^5}\left\{-2\frac{\dot a}{a} D_{P,i}+2\dot D_{P,i}\right.\nonumber\\&&\left.+
\frac{1}{2a}\left[4a\left(\dot \phi_G \phi_{G,i}+2\phi_G\dot{\phi}_{G,i}   \right)+8\dot a \phi_G\phi_{G,i}+2\omega_{k,i}\phi_{G,k}-\omega_{i}\nabla^2 \phi_G-2\omega_{k}\phi_{G,ki}\right]\right\}\nonumber\\&&=-\frac{1}{c^3}8\pi Ga\bar\rho(1+\delta)  v^*_i-\frac{1}{c^5}8\pi Ga\bar\rho(1+\delta) \left[v^*_i\left(v^{*2}-4\phi_G\right)\right]\;.
\end{eqnarray}
\end{subequations}

The continuity equation (\ref{continuity}) and the Euler equation (\ref{eulero}) become

\begin{eqnarray}
\label{contstar}
&&\frac{d\delta}{dt}+\frac{v^{*i}_{\;\;\;,i}}{a}(\delta+1)-\frac{1}{c^2}\left[(\delta+1)\left(3\frac{d\phi_G}{dt}+\frac{v^{*}_k\phi_{G,k}}{a}+\frac{\dot a}{a}v^{*2}\right) -\frac{1}{a}\omega^j \delta_{,j}\right]=0\;.\\
\label{eulerostar}
&&\frac{dv^*_i}{dt}+\frac{\dot a}{a} v^*_i+\frac{1}{a}\phi_{G,i}+\frac{1}{c^2}\left[\frac{1}{a}\phi_{G_,i}(4\phi_G+v^{*2})-3v^*_i\frac{d\phi_G}{dt}+\frac{1}{a}D_{P,i}-\frac{1}{a}v^*_iv^*_j\phi_{G}^{\phantom{G},j}-\frac{\dot a}{a}v^{*2}v^*_i+\frac{1}{a}\omega_{j,i}v^{*j}   +\frac{1}{a}\omega^j v^{*i}_{,j}  \right]=0\;,\nonumber\\
\end{eqnarray}
where here and in the following the convective derivative of a quantity $Q$ is defined in terms of $v^*$: 
\begin{equation}
\frac{dQ}{dt}=\dot Q +\frac{v_*^iQ_{,i}}{a}\text{.}
\end{equation}
\subsection{Constraint type and evolution type equations}

Starting from the new definitions given above, we want to obtain a final set of equations,   providing a unified  description of  gravitational instability in both  the linear and the nonlinear regimes, 
valid for 
fluctuations on all scales.
  
From Eqs.(\ref{psp}) and (\ref{trp}) we obtain an equation involving only the potential $\phi_G$, sourced by the matter and velocity perturbations: 
\begin{eqnarray}\label{eqphiG}
&&\frac{1}{c^2}\frac{2}{3}\nabla^2\phi_G+\frac{1}{c^4}\left[a^2\left(\ddot\phi_G+2\frac{\dot a}{a}\dot \phi_G+2\frac{\ddot a}{a}\phi_G-\left(\frac{\dot a}{a}\right)^2\phi_G\right)+\frac{2}{3}\nabla^2\phi_G^2-\frac{3}{2}\phi_{G,i}\phi_{G}^{\;\;,i}\right]=\frac{1}{c^2}\frac{8\pi G}{3}a^2\bar\rho\delta \nonumber \\ 
&&+\frac{1}{c^4}4\pi G a^2\bar\rho(1+\delta)v^{*2}\;.
\end{eqnarray}
Note that the linearized version of this equation is  a combination of Eqs.\ (\ref{pslinp}) and (\ref{trlinnew}).

A constraint equation for $D_{P}$ is obtained from Eq.~(\ref{scalargij}) [or, equivalently, by applying the operator $\partial_j\partial^i$ on both sides of Eq.\ (\ref{tfbp})]:
\begin{eqnarray}
\label{constraint2}
\frac{1}{c^4}\frac{2}{3}\nabla^2\nabla^2 D_{P}&=&- \frac{1}{c^4}\left[\left(\phi_{G,i}\phi_G^{\phantom{G},j}\right)_{,j}^{\;\;,i}-\frac{1}{3}\nabla^2\left(\phi_{G,i}\phi_{G}^{\phantom{G},i}\right)\right]-\frac{1}{c^4}4\pi G a^2\bar\rho\left[(1+\delta)\left(v^*_iv^{*j}-\frac{1}{3}v^{*2}\,\delta_i^j\right)\right]_{,j}^{,i}\;.
\end{eqnarray}
Rewriting the definitions of $\mathcal{A}_i^j$ and $\mathcal{S}_i^j$, Eqs.\  (\ref{A}) and (\ref{B}), in terms of $\phi_G$ and $v^{*i}$, also neglecting a higher-order contribution for $D_P$,
we obtain
\begin{eqnarray}
&&\mathcal{A}_i^j=2\phi_{G,i}\phi_G^{\;\;,j}-\frac{2}{3}\delta_i^j\phi_{G,k} \phi_{G}^{\;\;,k}\;,\\
&&\mathcal{S}_i^j=(1+\delta)(v_i^*v^{*j}-\frac{1}{3}v^{*2}\,\delta_i^j)\;.
\end{eqnarray}
With  these definitions Eq.\ (\ref{constraint2}) becomes
\begin{eqnarray}
\label{constraint3}
\frac{1}{c^4}\frac{4}{3}\nabla^2\nabla^2 D_{P}&=&- \frac{1}{c^4} \mathcal{A}_{k,j}^{j,k}-\frac{1}{c^4}8\pi G a^2\bar\rho\mathcal{S}_{k,j}^{j,k}\;.
\end{eqnarray}

Finally, let us rewrite Eq. (\ref{dynvects2}) as an evolution equation for the frame-dragging potential $\omega_i$
\begin{eqnarray}
\label{dynvects3}
&&\frac{1}{c^4}\nabla^2\nabla^2\left(\frac{1}{2a}\dot \omega_{i} +\frac{1}{a}\frac{\dot a}{a} \omega_{i}\right) = \frac{1}{c^4}\left[\frac{1}{a^2}\left(\nabla^2\mathcal{A}_{i,j}^{j}-\mathcal{A}_{k,ji}^{i,k}\right)+8\pi G \bar\rho \left(\nabla^2\mathcal{S}_{i,j}^{j}-\mathcal{S}_{k,ji}^{i,k}\right)\right]\;,
\end{eqnarray}
while the constraint Eq.\ (\ref{hij}) for $h_{ij}$ remains formally unchanged,  
\begin{eqnarray}
\label{hij3}
&&\frac{1}{c^4}\nabla^2\nabla^2\nabla^2h_{i}^{j}=\frac{1}{c^4}\left[-\mathcal{A}_{k,li}^{l,kj}-\nabla^2\mathcal{A}_{k,l}^{l,k}\delta_{i}^{j}+2\nabla^2\mathcal{A}_{i,k}^{k,j}   +2\nabla^2\mathcal{A}^{k}_{l,ki} \delta^{lj}-2\nabla^2\nabla^2\mathcal{A}_i^j\right.\nonumber\\
&&\left.+8\pi G a^2 \bar\rho\left(-\mathcal{S}_{k,li}^{l,kj}-\nabla^2\mathcal{S}_{k,l}^{l,k}\delta_i^j+2\nabla^2\mathcal{S}_{l,ki}^{k} \delta^{lj}+2\nabla^2\mathcal{S}^{k,j}_{i,k}-2\nabla^2\nabla^2\mathcal{S}_i^j\right)\right]\;.
\end{eqnarray}

At  1PF  order the  variables $h_{ij}$ and $D_P$ are not dynamical, i.e.\ they do not satisfy an evolution equation, rather they are given in terms of other variables by constraint equations. We can, therefore, conclude that the 1PF correction to Newtonian gravity introduces three nondynamical geometrical degrees of freedom (two d.o.f. in $h_{ij}$ and 1 d.o.f. in $D_P$) and provides the evolution equation for the gravitational potential $\phi_G$ and the frame dragging vector  potential $\omega_i$. 
For comparison, at 0PF Newtonian order we only have three nondynamical d.o.f.: one in the scalar gravitational potential and two in the gravitomagnetic vector potential. In full relativistic theory we would have six dynamical d.o.f., as is already apparent in first-order perturbation theory. Therefore, in order to recover a fully dynamical theory we should extend our post-Friedmann scheme to the 2PF order, cf.\   \cite{Szekeres:1999ny}.
However, the equations above, given that they extend the fully nonlinear 0PF Newtonian equations and (as shown in Sec \ref{linear_limit}) include the scalar and vector first-order perturbation equations,  are sufficient to study structure formation at all scales.

\subsection{New matter variables and simplified conservation equations}

There exists a further change of variable in the matter density, such that the 1PF continuity equation formally takes the usual Newtonian form \cite{Chandrasekhar:1965ch}.
 
Defining 
\begin{equation}
\hat\rho:=a^{-3}\rho(-g)^{1/2}u^0=\rho\left[1+\frac{1}{c^2}\left(\frac{1}{2}v^{*2}-3\phi_G\right)\right]\;,
\end{equation}
 and $\hat\delta:=(\hat\rho-\bar\rho)/\bar\rho$, equation (\ref{contstar}) becomes
\begin{eqnarray}
\label{ari}
\frac{d\hat\delta}{dt}+\frac{v^{*i}_{\;\;\;,i}}{a}(\hat\delta+1)=0\;;
\end{eqnarray}
therefore ``the mass defined in terms of the density $\hat\rho$ is conserved" \cite{Chandrasekhar:1965ch} in the usual Newtonian sense\footnote{Of course, mass is covariantly conserved in GR \cite{ellis2012relativistic,ellis1971Varenna}. For a discussion in second-order perturbation theory see \cite{Bertacca:2015mca}.}. Similarly, we can introduce a new velocity field $\hat v_i$, such that the 1PF Euler equation simplifies. Defining 
\begin{eqnarray}
\hat{v_i}=v^*_i\left[1+\frac{1}{c^2}\left(\frac{1}{2}v^{*2}-3\phi_G\right)\right]\;,
\end{eqnarray}
equation (\ref{eulerostar}) takes the form
\begin{eqnarray}
\frac{d\hat{v}_i}{dt}+\frac{\dot a}{a} \hat v_i+\frac{1}{a}\phi_{G,i}+\frac{1}{c^2}\left[\frac{\phi_{G_,i}}{a}\left(\phi_G+\frac{3}{2}\hat v^{2}\right)+\frac{D_{P_,i}}{a}+\frac{\omega_{j,i}\hat v^{j} }{a}\right]=0\;.
\end{eqnarray}
Let us note, however, that using this definition of velocity in the continuity equation (\ref{ari}) we would get some extra terms.


\section{Conclusions}\label{Conclusions}

Nonlinear structure formation in cosmology is traditionally studied with Newtonian N-body simulations and various approximation methods, e.g. Lagrangian perturbation theory. Relativistic perturbation theory and other approximations such as a gradient expansion are used to study small fluctuations in the early Universe, in the CMB and on very large scales in the matter era. Thus there is a gap between methods used to study large and small scales. 
Various authors \cite{Chisari:2011iq,Green:2011wc,Bar05,Bar10,Bruni:2011ta,Bruni:2013mua,Kopp:2013tqa,Bruni:2013qta,Bruni:2014xma,Adamek:2014xba,Green:2014aga,2014CQGra..31w4005V,Rigopoulos:2014rqa,2015arXiv150100799T,Yoo:2014kpa}
have recently pointed out that, given the high precision of current and future galaxy surveys, it is timely to investigate  possible GR effects on structure formation at all scales.

Assuming a standard flat $\Lambda$CDM cosmology, and a fluid approximation, 
in this Paper (Sec. \ref{NonlinearPN})  we have developed a unified  resummed nonlinear post-Friedmann formalism to study 
structure formation in the Universe. This relativistic scheme reduces to fully nonlinear Newtonian cosmology at leading order (Sec. \ref{Newtonianregime}) and, if linearized (Sec. \ref{linear_limit}),  to standard first-order relativistic perturbation theory.  
Thus our post-Friedmann formalism is valid on all scales, 
 bridging the existing gap in current studies of nonlinear structure formation.

We have focused on obtaining a set of approximate nonlinear equation, consistently  using the complete set of Einstein equations and the conservation equations.
Rather than using the standard iterative procedure of the perturbative analysis where the equations are derived and solved 
order by order, we have constructed  a resummed approximation scheme, which includes 
the first relativistic corrections to the equations of Newtonian cosmology, where a set of appropriately defined resummed variables satisfies a set of nonlinear equations.

Our equations could be used to implement GR corrections in N-body simulations. This would be important 
in order to take into account causal, retardation and other GR effects that may be non-negligible for simulations aiming at a 1\%  accuracy \cite{Laureijs:2011gra} 
 on scales of the order of the Hubble horizon. 
Indeed, we should consider  that the relevant fluctuations for the formation of large-scale structure have not 
 always been much smaller than the Hubble scale in the past. For instance, the present physical length of the horizon scale at  decoupling  is of the order of $ct_{dec}(1+z_{dec})\approx80h^{-1}$Mpc. 
Since galaxy surveys are going to cover a large proportion of the Hubble volume and aim at high precision measurements  \cite{Laureijs:2011gra}, we should consider 
GR corrections for the large and intermediate scales where the Newtonian approximation is not good enough. A first step in this direction has been done in \cite{Bruni:2013mua,2015arXiv150100799T}, computing the frame-dragging gravitomagnetic vector potentials from N-body Newtonian simulations, and in \cite{Thomas:2014aga}, where weak-lensing has been considered in the post-Friedmann framework. Our equations could also be used to include GR corrections in approximate Newtonian studies of various nonlinear effects such as BAO and CDM halos \cite{Crocce:2007dt,Matsubara:2008wx,Taruya:2010mx,Wang:2014gda,McCullagh:2014jsa,Seljak:2015rea}.
A numerical implementation of our resummed equations 
is now  a timely as well as doable goal: codes going beyond the quasistatic approximation have now been successfully developed for modified gravity \cite{Llinares:2013jua}, and codes 
specifically including GR corrections 
are in the making \cite{Adamek:2014xba}.
In this paper we have provided the theoretical framework aimed at this goal. 


\section*{Acknowledgements} 
This work is largely based on I.M. Ph.D thesis \cite{milillo:2010}.
We thank Daniel B.\ Thomas, Cornelius Rampf and Eleonora Villa for useful discussions and comments. 
During the preparation of this work D.B. was supported by the Deutsche Forschungsgemeinschaft through the Transregio 33, The Dark Universe, by the South African Square Kilometre Array Project and by  a Royal Society (UK)/ NRF (SA) exchange grant. M.B. was supported by the STFC Grants N0. ST/H002774/1, No. ST/K00090X/1 and No. ST/L005573/1. 
A.M. is supported by the NSF grants No. 1205864, No. 1212433 and No. 1333360.

\appendix

\section{Some Useful Quantities}
\label{Appendix}
\begin{subequations}
In this Appendix, for completeness  we collect some useful expression for the energy-momentum tensor and some geometrical quantities.
\begin{eqnarray}
T_{00}&=&\rho c^2+\rho(v^2-2U_N)+\frac{\rho}{c^2}\left[v^4-4U_P+2v^2 V_N+2U_N^{\phantom{N}2}\right]\\
T_{0i}&=&-\rho a cv_i+\frac{\rho a}{c}[B^N_{\phantom{N}i}-v_i(v^2+2V_N)]\\
T_{ij}&=&\rho a^2v_iv_j+\frac{a^2 \rho}{c^2}\left[(4V_N+2U_N+v^2)v_iv_j- 2B^{N}_{\phantom{N}(i} v_{j)}\right]
\end{eqnarray}
\end{subequations}
Riemann Tensor:
\begin{subequations}
\begin{eqnarray}
R_{i00j} & = & \frac{1}{c^2}[a\ddot a\delta_{ij}+U_{N,ij}]+\frac{1}{c^4}\left[a^2\delta_{ij}\left(\frac{\dot a}{a}\dot U_N+\ddot V_N+2\frac{\dot a}{a}
\dot V_N+2\frac{\ddot a}{a}V_N+U_{N,k}V_{N}^{\phantom{N},k}\right)\right. \nonumber\\
&+&a\dot B^N_{\phantom{N}(i,j)}+\dot a B^N_{\phantom{N}(i,j)}+2 U_{P,ij}-\left.U_{N,i}U_{N,j}-2U_{N,(i}V_{N,j)}-2U_NU_{N,ij}\right]\\
R_{0ijk} & = & \frac{1}{c^3}\left[2a^2\left(\dot V_{N,[k}+\frac{\dot a}{a}U_{N,[k}\right)\delta_{j]i}+aB^N_{\phantom{N}[j,k]i}\right]\\
R_{ijkl} & = & \frac{1}{c^2}\left[2a^2\dot a^2\delta_{i[k}\delta_{l]j}+2a^2(V_{N,j[k}\delta_{l]i}-V_{N,i[k}\delta_{l]j}) \right]+\frac{1}{c^4}\left[\delta_{i[l}\delta_{k]j}a^22\left(-4\dot a^2V_N-2a\dot a\dot V_N-2\dot a^2 U_N+V_{N,n}V_{N,n}\right)\right. \nonumber\\
 & + & 2a^2\left(\frac{1}{2}\dot aB^{N}_{\phantom{N}i,[k}+\frac{1}{2}\dot aB^N_{\phantom{N}[k,|i|}+V_{N,i}V_{N,[k}-2V_{N,i[k}V_N-2V_{P,i[k}\right)\delta_{l]j}-2a^2\left(\frac{1}{2}\dot aB^N_{\phantom{N}j,[k}+\frac{1}{2}\dot aB^N_{\phantom{N}[k,|j|}+V_{N,j}V_{N,[k}\right. \nonumber\\
 & - &\left. \left.2V_{N,j[k}V-2V_{B,j[k}\right)\delta_{l]i}\right]
\end{eqnarray}
\end{subequations}
Ricci Tensor:
\begin{subequations}
\begin{eqnarray}
R_{00} &=&-\frac{1}{c^2a^2}(\nabla^2U_N+3a\ddot a )-\frac{1}{c^4a^2}\left[V_{N,k}U_{N,k}-2\nabla^2U_N(U_N+V_N)-U_{N,k}U_N^{\phantom{N},k}+2\nabla^2U_P+3a^2\ddot V_N+6a\dot a\dot V_N+3a\dot a \dot U_N\right]\nonumber\\ \\
R_{0i}&=&-\frac{1}{2c^3a}(4\dot a U_{N,i}+4a\dot V_{N,i}-\nabla^2 B^N_{\phantom{N}i}) \\
R_{ij}&=&\frac{1}{c^2}\left[\left(a\ddot a+2\dot a^2-\nabla^2 V_N \right)\delta_{ij}+(U_N-V_N)_{,ij}\right]+\frac{1}{c^4}\left[2\dot a B^N_{\phantom{N}(i,j)}+
a\dot B^{N}_{\phantom{N}(i,j)}+V_{N,i}V_{N,j}-U_{N,i}U_{N,j}-2V_{N,(i}U_{N,j)} \right.\nonumber\\
& + &2(U_P-V_P)_{,ij}+  \delta_{ij}\left(V_{N,k}U_N^{\phantom{N},k}-V_{N,k}V_N^{\phantom{N},k}+4\dot a^2(V_N+U_N)-2\nabla^2V_P+6a \dot a\dot V_N \right. \nonumber\\
&+&\left. \left.2a\ddot a(U_N+ V_N)+a^2\ddot V_N+ a\dot a\dot U_N-\frac{1}{2}\nabla^2h_{ij}\right)\right].
\end{eqnarray}
\end{subequations}
Ricci scalar:
\begin{eqnarray}
R&=&\frac{1}{c^2}\left[6\left(\frac{\ddot a}{a}+\left(\frac{\dot a}{a}\right)^2\right)+2\frac{\nabla^2}{a^2}(U_N-2V_N)\right]+\frac{1}{c^4}\left\{6\ddot V_N+6\frac{\dot a}{a}\left(\dot U_N+4\dot V_N\right)\right.\nonumber\\&-&\left.\frac{2}{a^2}\left(V_{N,i}-U_{N,i}\right)^2-2V\frac{\nabla^2}{a^2}(U_N-2V_N)+\frac{2}{a^2}V_{N,i}U_{N,i}+4\frac{\nabla^2}{a^2}(U_P-2V_P)\right\}
\end{eqnarray}

\bibliography{articolibib}

\begin{thebibliography}{100}
\expandafter\ifx\csname natexlab\endcsname\relax\def\natexlab#1{#1}\fi
\expandafter\ifx\csname bibnamefont\endcsname\relax
  \def\bibnamefont#1{#1}\fi
\expandafter\ifx\csname bibfnamefont\endcsname\relax
  \def\bibfnamefont#1{#1}\fi
\expandafter\ifx\csname citenamefont\endcsname\relax
  \def\citenamefont#1{#1}\fi
\expandafter\ifx\csname url\endcsname\relax
  \def\url#1{\texttt{#1}}\fi
\expandafter\ifx\csname urlprefix\endcsname\relax\def\urlprefix{URL }\fi
\providecommand{\bibinfo}[2]{#2}
\providecommand{\eprint}[2][]{\url{#2}}

\bibitem[{\citenamefont{Bruni et~al.}(2014{\natexlab{a}})\citenamefont{Bruni,
  Thomas, and Wands}}]{Bruni:2013mua}
\bibinfo{author}{\bibfnamefont{M.}~\bibnamefont{Bruni}},
  \bibinfo{author}{\bibfnamefont{D.~B.} \bibnamefont{Thomas}},
  \bibnamefont{and} \bibinfo{author}{\bibfnamefont{D.}~\bibnamefont{Wands}},
  \bibinfo{journal}{Phys.Rev.} \textbf{\bibinfo{volume}{D89}},
  \bibinfo{pages}{044010} (\bibinfo{year}{2014}{\natexlab{a}}),
  \eprint{1306.1562}.

\bibitem[{\citenamefont{{Thomas} et~al.}(2015)\citenamefont{{Thomas}, {Bruni},
  and {Wands}}}]{2015arXiv150100799T}
\bibinfo{author}{\bibfnamefont{D.~B.} \bibnamefont{{Thomas}}},
  \bibinfo{author}{\bibfnamefont{M.}~\bibnamefont{{Bruni}}}, \bibnamefont{and}
  \bibinfo{author}{\bibfnamefont{D.}~\bibnamefont{{Wands}}},
  \bibinfo{journal}{ArXiv e-prints}  (\bibinfo{year}{2015}),
  \eprint{1501.00799}.

\bibitem[{\citenamefont{Peebles}(1984)}]{Peebles:1984ge}
\bibinfo{author}{\bibfnamefont{P.~J.~E.} \bibnamefont{Peebles}},
  \bibinfo{journal}{Astrophys. J.} \textbf{\bibinfo{volume}{284}},
  \bibinfo{pages}{439} (\bibinfo{year}{1984}).

\bibitem[{\citenamefont{Efstathiou et~al.}(1990)\citenamefont{Efstathiou,
  Sutherland, and Maddox}}]{Efstathiou:1990xe}
\bibinfo{author}{\bibfnamefont{G.}~\bibnamefont{Efstathiou}},
  \bibinfo{author}{\bibfnamefont{W.~J.} \bibnamefont{Sutherland}},
  \bibnamefont{and} \bibinfo{author}{\bibfnamefont{S.~J.}
  \bibnamefont{Maddox}}, \bibinfo{journal}{Nature}
  \textbf{\bibinfo{volume}{348}}, \bibinfo{pages}{705} (\bibinfo{year}{1990}).

\bibitem[{\citenamefont{Tegmark et~al.}(2004)}]{Tegmark:2003ud}
\bibinfo{author}{\bibfnamefont{M.}~\bibnamefont{Tegmark}} \bibnamefont{et~al.}
  (\bibinfo{collaboration}{SDSS}), \bibinfo{journal}{Phys. Rev.}
  \textbf{\bibinfo{volume}{D69}}, \bibinfo{pages}{103501}
  (\bibinfo{year}{2004}), \eprint{astro-ph/0310723}.

\bibitem[{\citenamefont{Ellis et~al.}(2012)\citenamefont{Ellis, Maartens, and
  MacCallum}}]{ellis2012relativistic}
\bibinfo{author}{\bibfnamefont{G.~F.} \bibnamefont{Ellis}},
  \bibinfo{author}{\bibfnamefont{R.}~\bibnamefont{Maartens}}, \bibnamefont{and}
  \bibinfo{author}{\bibfnamefont{M.~A.} \bibnamefont{MacCallum}},
  \emph{\bibinfo{title}{Relativistic cosmology}} (\bibinfo{publisher}{Cambridge
  University Press}, \bibinfo{year}{2012}).

\bibitem[{\citenamefont{Maartens}(2011)}]{Maartens:2011yx}
\bibinfo{author}{\bibfnamefont{R.}~\bibnamefont{Maartens}},
  \bibinfo{journal}{Phil.Trans.Roy.Soc.Lond.} \textbf{\bibinfo{volume}{A369}},
  \bibinfo{pages}{5115} (\bibinfo{year}{2011}), \eprint{1104.1300}.

\bibitem[{\citenamefont{de~Bernardis et~al.}(2000)}]{deBernardis:2000gy}
\bibinfo{author}{\bibfnamefont{P.}~\bibnamefont{de~Bernardis}}
  \bibnamefont{et~al.} (\bibinfo{collaboration}{Boomerang Collaboration}),
  \bibinfo{journal}{Nature} \textbf{\bibinfo{volume}{404}},
  \bibinfo{pages}{955} (\bibinfo{year}{2000}), \eprint{astro-ph/0004404}.

\bibitem[{\citenamefont{Ade et~al.}(2014)}]{Ade:2013zuv}
\bibinfo{author}{\bibfnamefont{P.}~\bibnamefont{Ade}} \bibnamefont{et~al.}
  (\bibinfo{collaboration}{Planck}), \bibinfo{journal}{Astron.Astrophys.}
  \textbf{\bibinfo{volume}{571}}, \bibinfo{pages}{A16} (\bibinfo{year}{2014}),
  \eprint{1303.5076}.

\bibitem[{\citenamefont{Aubourg et~al.}(2014)\citenamefont{Aubourg, Bailey,
  Bautista, Beutler, Bhardwaj et~al.}}]{Aubourg:2014yra}
\bibinfo{author}{\bibfnamefont{E.}~\bibnamefont{Aubourg}},
  \bibinfo{author}{\bibfnamefont{S.}~\bibnamefont{Bailey}},
  \bibinfo{author}{\bibfnamefont{J.~E.} \bibnamefont{Bautista}},
  \bibinfo{author}{\bibfnamefont{F.}~\bibnamefont{Beutler}},
  \bibinfo{author}{\bibfnamefont{V.}~\bibnamefont{Bhardwaj}},
  \bibnamefont{et~al.} (\bibinfo{year}{2014}), \eprint{1411.1074}.

\bibitem[{\citenamefont{Bertschinger}(1996)}]{Bertschinger:1993xt}
\bibinfo{author}{\bibfnamefont{E.}~\bibnamefont{Bertschinger}},
  \emph{\bibinfo{title}{\normalfont Cosmological dynamics, in \itshape
  Cosmology and Large Scale Structure, proc. Les Houches Summer School, Session
  LX}} (\bibinfo{publisher}{ed.\ R. Schaeffer, J. Silk, M. Spiro, and J.
  Zinn-Justin , Amsterdam: Elsevier Science.}, \bibinfo{year}{1996}),
  \eprint{astro-ph/9503125}.

\bibitem[{\citenamefont{Malik and Wands}(2009)}]{Malik:2008im}
\bibinfo{author}{\bibfnamefont{K.~A.} \bibnamefont{Malik}} \bibnamefont{and}
  \bibinfo{author}{\bibfnamefont{D.}~\bibnamefont{Wands}},
  \bibinfo{journal}{Phys.Rept.} \textbf{\bibinfo{volume}{475}},
  \bibinfo{pages}{1} (\bibinfo{year}{2009}), \eprint{0809.4944}.

\bibitem[{\citenamefont{Bertschinger}(1998)}]{Bertschinger:1998tv}
\bibinfo{author}{\bibfnamefont{E.}~\bibnamefont{Bertschinger}},
  \bibinfo{journal}{Ann. Rev. Astron. Astrophys.}
  \textbf{\bibinfo{volume}{36}}, \bibinfo{pages}{599} (\bibinfo{year}{1998}).

\bibitem[{\citenamefont{Ehlers and Buchert}(1997)}]{Ehlers:1996wg}
\bibinfo{author}{\bibfnamefont{J.}~\bibnamefont{Ehlers}} \bibnamefont{and}
  \bibinfo{author}{\bibfnamefont{T.}~\bibnamefont{Buchert}},
  \bibinfo{journal}{Gen.Rel.Grav.} \textbf{\bibinfo{volume}{29}},
  \bibinfo{pages}{733} (\bibinfo{year}{1997}), \eprint{astro-ph/9609036}.

\bibitem[{\citenamefont{Bernardeau et~al.}(2002)\citenamefont{Bernardeau,
  Colombi, Gaztanaga, and Scoccimarro}}]{Bernardeau:2001qr}
\bibinfo{author}{\bibfnamefont{F.}~\bibnamefont{Bernardeau}},
  \bibinfo{author}{\bibfnamefont{S.}~\bibnamefont{Colombi}},
  \bibinfo{author}{\bibfnamefont{E.}~\bibnamefont{Gaztanaga}},
  \bibnamefont{and}
  \bibinfo{author}{\bibfnamefont{R.}~\bibnamefont{Scoccimarro}},
  \bibinfo{journal}{Phys.Rept.} \textbf{\bibinfo{volume}{367}},
  \bibinfo{pages}{1} (\bibinfo{year}{2002}), \eprint{astro-ph/0112551}.

\bibitem[{\citenamefont{Crocce and Scoccimarro}(2008)}]{Crocce:2007dt}
\bibinfo{author}{\bibfnamefont{M.}~\bibnamefont{Crocce}} \bibnamefont{and}
  \bibinfo{author}{\bibfnamefont{R.}~\bibnamefont{Scoccimarro}},
  \bibinfo{journal}{Phys.Rev.} \textbf{\bibinfo{volume}{D77}},
  \bibinfo{pages}{023533} (\bibinfo{year}{2008}), \eprint{0704.2783}.

\bibitem[{\citenamefont{Matsubara}(2008)}]{Matsubara:2008wx}
\bibinfo{author}{\bibfnamefont{T.}~\bibnamefont{Matsubara}},
  \bibinfo{journal}{Phys.Rev.} \textbf{\bibinfo{volume}{D78}},
  \bibinfo{pages}{083519} (\bibinfo{year}{2008}), \eprint{0807.1733}.

\bibitem[{\citenamefont{Taruya et~al.}(2010)\citenamefont{Taruya, Nishimichi,
  and Saito}}]{Taruya:2010mx}
\bibinfo{author}{\bibfnamefont{A.}~\bibnamefont{Taruya}},
  \bibinfo{author}{\bibfnamefont{T.}~\bibnamefont{Nishimichi}},
  \bibnamefont{and} \bibinfo{author}{\bibfnamefont{S.}~\bibnamefont{Saito}},
  \bibinfo{journal}{Phys.Rev.} \textbf{\bibinfo{volume}{D82}},
  \bibinfo{pages}{063522} (\bibinfo{year}{2010}), \eprint{1006.0699}.

\bibitem[{\citenamefont{Wang and Szalay}(2014)}]{Wang:2014gda}
\bibinfo{author}{\bibfnamefont{X.}~\bibnamefont{Wang}} \bibnamefont{and}
  \bibinfo{author}{\bibfnamefont{A.}~\bibnamefont{Szalay}}
  (\bibinfo{year}{2014}), \eprint{1411.4117}.

\bibitem[{\citenamefont{McCullagh and Szalay}(2015)}]{McCullagh:2014jsa}
\bibinfo{author}{\bibfnamefont{N.}~\bibnamefont{McCullagh}} \bibnamefont{and}
  \bibinfo{author}{\bibfnamefont{A.~S.} \bibnamefont{Szalay}},
  \bibinfo{journal}{Astrophys.J.} \textbf{\bibinfo{volume}{798}},
  \bibinfo{pages}{137} (\bibinfo{year}{2015}), \eprint{1411.1249}.

\bibitem[{\citenamefont{Seljak and Vlah}(2015)}]{Seljak:2015rea}
\bibinfo{author}{\bibfnamefont{U.}~\bibnamefont{Seljak}} \bibnamefont{and}
  \bibinfo{author}{\bibfnamefont{Z.}~\bibnamefont{Vlah}}
  (\bibinfo{year}{2015}), \eprint{1501.07512}.

\bibitem[{\citenamefont{Ade et~al.}(2015)}]{Planck:2015xua}
\bibinfo{author}{\bibfnamefont{P.}~\bibnamefont{Ade}} \bibnamefont{et~al.}
  (\bibinfo{collaboration}{Planck Collaboration}) (\bibinfo{year}{2015}),
  \eprint{1502.01589}.

\bibitem[{\citenamefont{Betoule et~al.}(2014)}]{Betoule:2014frx}
\bibinfo{author}{\bibfnamefont{M.}~\bibnamefont{Betoule}} \bibnamefont{et~al.}
  (\bibinfo{collaboration}{SDSS Collaboration}),
  \bibinfo{journal}{Astron.Astrophys.} \textbf{\bibinfo{volume}{568}},
  \bibinfo{pages}{A22} (\bibinfo{year}{2014}), \eprint{1401.4064}.

\bibitem[{\citenamefont{Riess et~al.}(2011)\citenamefont{Riess, Macri,
  Casertano, Lampeitl, Ferguson et~al.}}]{Riess:2011yx}
\bibinfo{author}{\bibfnamefont{A.~G.} \bibnamefont{Riess}},
  \bibinfo{author}{\bibfnamefont{L.}~\bibnamefont{Macri}},
  \bibinfo{author}{\bibfnamefont{S.}~\bibnamefont{Casertano}},
  \bibinfo{author}{\bibfnamefont{H.}~\bibnamefont{Lampeitl}},
  \bibinfo{author}{\bibfnamefont{H.~C.} \bibnamefont{Ferguson}},
  \bibnamefont{et~al.}, \bibinfo{journal}{Astrophys.J.}
  \textbf{\bibinfo{volume}{730}}, \bibinfo{pages}{119} (\bibinfo{year}{2011}),
  \eprint{1103.2976}.

\bibitem[{\citenamefont{{Eisenstein} et~al.}(2011)\citenamefont{{Eisenstein},
  {Weinberg}, {Agol}, {Aihara}, {Allende Prieto}, {Anderson}, {Arns},
  {Aubourg}, {Bailey}, {Balbinot} et~al.}}]{SDSS}
\bibinfo{author}{\bibfnamefont{D.~J.} \bibnamefont{{Eisenstein}}},
  \bibinfo{author}{\bibfnamefont{D.~H.} \bibnamefont{{Weinberg}}},
  \bibinfo{author}{\bibfnamefont{E.}~\bibnamefont{{Agol}}},
  \bibinfo{author}{\bibfnamefont{H.}~\bibnamefont{{Aihara}}},
  \bibinfo{author}{\bibfnamefont{C.}~\bibnamefont{{Allende Prieto}}},
  \bibinfo{author}{\bibfnamefont{S.~F.} \bibnamefont{{Anderson}}},
  \bibinfo{author}{\bibfnamefont{J.~A.} \bibnamefont{{Arns}}},
  \bibinfo{author}{\bibfnamefont{{\'E}.}~\bibnamefont{{Aubourg}}},
  \bibinfo{author}{\bibfnamefont{S.}~\bibnamefont{{Bailey}}},
  \bibinfo{author}{\bibfnamefont{E.}~\bibnamefont{{Balbinot}}},
  \bibnamefont{et~al.}, \bibinfo{journal}{Astron.J.}
  \textbf{\bibinfo{volume}{142}}, \bibinfo{eid}{72} (\bibinfo{year}{2011}),
  \eprint{1101.1529}.

\bibitem[{\citenamefont{Parkinson et~al.}(2012)\citenamefont{Parkinson,
  Riemer-Sorensen, Blake, Poole, Davis et~al.}}]{WiggleZ}
\bibinfo{author}{\bibfnamefont{D.}~\bibnamefont{Parkinson}},
  \bibinfo{author}{\bibfnamefont{S.}~\bibnamefont{Riemer-Sorensen}},
  \bibinfo{author}{\bibfnamefont{C.}~\bibnamefont{Blake}},
  \bibinfo{author}{\bibfnamefont{G.~B.} \bibnamefont{Poole}},
  \bibinfo{author}{\bibfnamefont{T.~M.} \bibnamefont{Davis}},
  \bibnamefont{et~al.}, \bibinfo{journal}{Phys.Rev.}
  \textbf{\bibinfo{volume}{D86}}, \bibinfo{pages}{103518}
  (\bibinfo{year}{2012}), \eprint{1210.2130}.

\bibitem[{\citenamefont{Song and Percival}(2009)}]{Song:2008qt}
\bibinfo{author}{\bibfnamefont{Y.-S.} \bibnamefont{Song}} \bibnamefont{and}
  \bibinfo{author}{\bibfnamefont{W.~J.} \bibnamefont{Percival}},
  \bibinfo{journal}{JCAP} \textbf{\bibinfo{volume}{0910}}, \bibinfo{pages}{004}
  (\bibinfo{year}{2009}), \eprint{0807.0810}.

\bibitem[{\citenamefont{Samushia et~al.}(2012)\citenamefont{Samushia, Percival,
  and Raccanelli}}]{Samushia:2011cs}
\bibinfo{author}{\bibfnamefont{L.}~\bibnamefont{Samushia}},
  \bibinfo{author}{\bibfnamefont{W.~J.} \bibnamefont{Percival}},
  \bibnamefont{and}
  \bibinfo{author}{\bibfnamefont{A.}~\bibnamefont{Raccanelli}},
  \bibinfo{journal}{Mon.Not.Roy.Astron.Soc.} \textbf{\bibinfo{volume}{420}},
  \bibinfo{pages}{2102} (\bibinfo{year}{2012}), \eprint{1102.1014}.

\bibitem[{\citenamefont{Macaulay et~al.}(2013)\citenamefont{Macaulay, Wehus,
  and Eriksen}}]{Macaulay:2013swa}
\bibinfo{author}{\bibfnamefont{E.}~\bibnamefont{Macaulay}},
  \bibinfo{author}{\bibfnamefont{I.~K.} \bibnamefont{Wehus}}, \bibnamefont{and}
  \bibinfo{author}{\bibfnamefont{H.~K.} \bibnamefont{Eriksen}},
  \bibinfo{journal}{Phys.Rev.Lett.} \textbf{\bibinfo{volume}{111}},
  \bibinfo{pages}{161301} (\bibinfo{year}{2013}), \eprint{1303.6583}.

\bibitem[{\citenamefont{Verde et~al.}(2013)\citenamefont{Verde, Protopapas, and
  Jimenez}}]{Verde:2013wza}
\bibinfo{author}{\bibfnamefont{L.}~\bibnamefont{Verde}},
  \bibinfo{author}{\bibfnamefont{P.}~\bibnamefont{Protopapas}},
  \bibnamefont{and} \bibinfo{author}{\bibfnamefont{R.}~\bibnamefont{Jimenez}},
  \bibinfo{journal}{Phys.Dark Univ.} \textbf{\bibinfo{volume}{2}},
  \bibinfo{pages}{166} (\bibinfo{year}{2013}), \eprint{1306.6766}.

\bibitem[{\citenamefont{Battye et~al.}(2014)\citenamefont{Battye, Charnock, and
  Moss}}]{Battye:2014qga}
\bibinfo{author}{\bibfnamefont{R.~A.} \bibnamefont{Battye}},
  \bibinfo{author}{\bibfnamefont{T.}~\bibnamefont{Charnock}}, \bibnamefont{and}
  \bibinfo{author}{\bibfnamefont{A.}~\bibnamefont{Moss}}
  (\bibinfo{year}{2014}), \eprint{1409.2769}.

\bibitem[{\citenamefont{Ruiz and Huterer}(2014)}]{Ruiz:2014hma}
\bibinfo{author}{\bibfnamefont{E.~J.} \bibnamefont{Ruiz}} \bibnamefont{and}
  \bibinfo{author}{\bibfnamefont{D.}~\bibnamefont{Huterer}}
  (\bibinfo{year}{2014}), \eprint{arXiv:1410.5832}.

\bibitem[{\citenamefont{Weinberg}(1989)}]{Weinberg:1988cp}
\bibinfo{author}{\bibfnamefont{S.}~\bibnamefont{Weinberg}},
  \bibinfo{journal}{Rev.Mod.Phys.} \textbf{\bibinfo{volume}{61}},
  \bibinfo{pages}{1} (\bibinfo{year}{1989}).

\bibitem[{\citenamefont{Amendola and Tsujikawa}(2010)}]{amendola2010dark}
\bibinfo{author}{\bibfnamefont{L.}~\bibnamefont{Amendola}} \bibnamefont{and}
  \bibinfo{author}{\bibfnamefont{S.}~\bibnamefont{Tsujikawa}},
  \emph{\bibinfo{title}{Dark energy: theory and observations}}
  (\bibinfo{publisher}{Cambridge University Press}, \bibinfo{year}{2010}).

\bibitem[{\citenamefont{Nojiri and Odintsov}(2006)}]{Nojiri:2006ri}
\bibinfo{author}{\bibfnamefont{S.}~\bibnamefont{Nojiri}} \bibnamefont{and}
  \bibinfo{author}{\bibfnamefont{S.~D.} \bibnamefont{Odintsov}},
  \bibinfo{journal}{eConf} \textbf{\bibinfo{volume}{C0602061}},
  \bibinfo{pages}{06} (\bibinfo{year}{2006}), \eprint{hep-th/0601213}.

\bibitem[{\citenamefont{Clifton et~al.}(2012)\citenamefont{Clifton, Ferreira,
  Padilla, and Skordis}}]{Clifton:2011jh}
\bibinfo{author}{\bibfnamefont{T.}~\bibnamefont{Clifton}},
  \bibinfo{author}{\bibfnamefont{P.~G.} \bibnamefont{Ferreira}},
  \bibinfo{author}{\bibfnamefont{A.}~\bibnamefont{Padilla}}, \bibnamefont{and}
  \bibinfo{author}{\bibfnamefont{C.}~\bibnamefont{Skordis}},
  \bibinfo{journal}{Phys.Rept.} \textbf{\bibinfo{volume}{513}},
  \bibinfo{pages}{1} (\bibinfo{year}{2012}), \eprint{1106.2476}.

\bibitem[{\citenamefont{Clarkson and Maartens}(2010)}]{Clarkson:2010uz}
\bibinfo{author}{\bibfnamefont{C.}~\bibnamefont{Clarkson}} \bibnamefont{and}
  \bibinfo{author}{\bibfnamefont{R.}~\bibnamefont{Maartens}},
  \bibinfo{journal}{Class.Quant.Grav.} \textbf{\bibinfo{volume}{27}},
  \bibinfo{pages}{124008} (\bibinfo{year}{2010}), \eprint{1005.2165}.

\bibitem[{\citenamefont{Buchert and R\"{a}s\"{a}nen}(2012)}]{Buchert:2011sx}
\bibinfo{author}{\bibfnamefont{T.}~\bibnamefont{Buchert}} \bibnamefont{and}
  \bibinfo{author}{\bibfnamefont{S.}~\bibnamefont{R\"{a}s\"{a}nen}},
  \bibinfo{journal}{Ann.Rev.Nucl.Part.Sci.} \textbf{\bibinfo{volume}{62}},
  \bibinfo{pages}{57} (\bibinfo{year}{2012}), \eprint{1112.5335}.

\bibitem[{\citenamefont{Clarkson et~al.}(2011)\citenamefont{Clarkson, Ellis,
  Larena, and Umeh}}]{Clarkson:2011zq}
\bibinfo{author}{\bibfnamefont{C.}~\bibnamefont{Clarkson}},
  \bibinfo{author}{\bibfnamefont{G.}~\bibnamefont{Ellis}},
  \bibinfo{author}{\bibfnamefont{J.}~\bibnamefont{Larena}}, \bibnamefont{and}
  \bibinfo{author}{\bibfnamefont{O.}~\bibnamefont{Umeh}},
  \bibinfo{journal}{Rept.Prog.Phys.} \textbf{\bibinfo{volume}{74}},
  \bibinfo{pages}{112901} (\bibinfo{year}{2011}), \eprint{1109.2314}.

\bibitem[{\citenamefont{Kolb et~al.}(2010)\citenamefont{Kolb, Marra, and
  Matarrese}}]{Kolb:2009rp}
\bibinfo{author}{\bibfnamefont{E.~W.} \bibnamefont{Kolb}},
  \bibinfo{author}{\bibfnamefont{V.}~\bibnamefont{Marra}}, \bibnamefont{and}
  \bibinfo{author}{\bibfnamefont{S.}~\bibnamefont{Matarrese}},
  \bibinfo{journal}{Gen.Rel.Grav.} \textbf{\bibinfo{volume}{42}},
  \bibinfo{pages}{1399} (\bibinfo{year}{2010}), \eprint{0901.4566}.

\bibitem[{\citenamefont{Anderson et~al.}(2014)}]{Anderson:2013zyy}
\bibinfo{author}{\bibfnamefont{L.}~\bibnamefont{Anderson}}
  \bibnamefont{et~al.}, \bibinfo{journal}{MNRAS}
  \textbf{\bibinfo{volume}{441}}, \bibinfo{pages}{24} (\bibinfo{year}{2014}),
  \eprint{1312.4877}.

\bibitem[{\citenamefont{Laureijs et~al.}(2011)}]{Laureijs:2011gra}
\bibinfo{author}{\bibfnamefont{R.}~\bibnamefont{Laureijs}} \bibnamefont{et~al.}
  (\bibinfo{collaboration}{EUCLID Collaboration}) (\bibinfo{year}{2011}),
  \eprint{1110.3193}.

\bibitem[{\citenamefont{Amendola et~al.}(2013)}]{Amendola:2012ys}
\bibinfo{author}{\bibfnamefont{L.}~\bibnamefont{Amendola}} \bibnamefont{et~al.}
  (\bibinfo{collaboration}{Euclid Theory Working Group}),
  \bibinfo{journal}{Living Rev.Rel.} \textbf{\bibinfo{volume}{16}},
  \bibinfo{pages}{6} (\bibinfo{year}{2013}), \eprint{1206.1225}.

\bibitem[{\citenamefont{Scaramella et~al.}(2015)}]{Scaramella:2015rra}
\bibinfo{author}{\bibfnamefont{R.}~\bibnamefont{Scaramella}}
  \bibnamefont{et~al.} (\bibinfo{collaboration}{Euclid Collaboration})
  (\bibinfo{year}{2015}), \eprint{1501.04908}.

\bibitem[{\citenamefont{Jarvis et~al.}(2015)\citenamefont{Jarvis, Bacon, Blake,
  Brown, Lindsay et~al.}}]{Jarvis:2015tqa}
\bibinfo{author}{\bibfnamefont{M.~J.} \bibnamefont{Jarvis}},
  \bibinfo{author}{\bibfnamefont{D.}~\bibnamefont{Bacon}},
  \bibinfo{author}{\bibfnamefont{C.}~\bibnamefont{Blake}},
  \bibinfo{author}{\bibfnamefont{M.~L.} \bibnamefont{Brown}},
  \bibinfo{author}{\bibfnamefont{S.~N.} \bibnamefont{Lindsay}},
  \bibnamefont{et~al.} (\bibinfo{year}{2015}), \eprint{1501.03825}.

\bibitem[{\citenamefont{{Schwarz} et~al.}(2015)\citenamefont{{Schwarz},
  {Bacon}, {Chen}, {Clarkson}, {Huterer}, {Kunz}, {Maartens}, {Raccanelli},
  {Rubart}, and {Starck}}}]{Schwarz:2015pqa}
\bibinfo{author}{\bibfnamefont{D.~J.} \bibnamefont{{Schwarz}}},
  \bibinfo{author}{\bibfnamefont{D.}~\bibnamefont{{Bacon}}},
  \bibinfo{author}{\bibfnamefont{S.}~\bibnamefont{{Chen}}},
  \bibinfo{author}{\bibfnamefont{C.}~\bibnamefont{{Clarkson}}},
  \bibinfo{author}{\bibfnamefont{D.}~\bibnamefont{{Huterer}}},
  \bibinfo{author}{\bibfnamefont{M.}~\bibnamefont{{Kunz}}},
  \bibinfo{author}{\bibfnamefont{R.}~\bibnamefont{{Maartens}}},
  \bibinfo{author}{\bibfnamefont{A.}~\bibnamefont{{Raccanelli}}},
  \bibinfo{author}{\bibfnamefont{M.}~\bibnamefont{{Rubart}}}, \bibnamefont{and}
  \bibinfo{author}{\bibfnamefont{J.-L.} \bibnamefont{{Starck}}},
  \bibinfo{journal}{ArXiv e-prints}  (\bibinfo{year}{2015}),
  \eprint{1501.03820}.

\bibitem[{\citenamefont{{Kitching} et~al.}(2015)\citenamefont{{Kitching},
  {Bacon}, {Brown}, {Bull}, {McEwen}, {Oguri}, {Scaramella}, {Takahashi}, {Wu},
  and {Yamauchi}}}]{Kitching:2015fra}
\bibinfo{author}{\bibfnamefont{T.~D.} \bibnamefont{{Kitching}}},
  \bibinfo{author}{\bibfnamefont{D.}~\bibnamefont{{Bacon}}},
  \bibinfo{author}{\bibfnamefont{M.~L.} \bibnamefont{{Brown}}},
  \bibinfo{author}{\bibfnamefont{P.}~\bibnamefont{{Bull}}},
  \bibinfo{author}{\bibfnamefont{J.~D.} \bibnamefont{{McEwen}}},
  \bibinfo{author}{\bibfnamefont{M.}~\bibnamefont{{Oguri}}},
  \bibinfo{author}{\bibfnamefont{R.}~\bibnamefont{{Scaramella}}},
  \bibinfo{author}{\bibfnamefont{K.}~\bibnamefont{{Takahashi}}},
  \bibinfo{author}{\bibfnamefont{K.}~\bibnamefont{{Wu}}}, \bibnamefont{and}
  \bibinfo{author}{\bibfnamefont{D.}~\bibnamefont{{Yamauchi}}},
  \bibinfo{journal}{ArXiv e-prints}  (\bibinfo{year}{2015}),
  \eprint{1501.03978}.

\bibitem[{\citenamefont{Chisari and Zaldarriaga}(2011)}]{Chisari:2011iq}
\bibinfo{author}{\bibfnamefont{N.~E.} \bibnamefont{Chisari}} \bibnamefont{and}
  \bibinfo{author}{\bibfnamefont{M.}~\bibnamefont{Zaldarriaga}},
  \bibinfo{journal}{Phys.Rev.} \textbf{\bibinfo{volume}{D83}},
  \bibinfo{pages}{123505} (\bibinfo{year}{2011}), \eprint{1101.3555}.

\bibitem[{\citenamefont{Green and Wald}(2012)}]{Green:2011wc}
\bibinfo{author}{\bibfnamefont{S.~R.} \bibnamefont{Green}} \bibnamefont{and}
  \bibinfo{author}{\bibfnamefont{R.~M.} \bibnamefont{Wald}},
  \bibinfo{journal}{Phys.Rev.} \textbf{\bibinfo{volume}{D85}},
  \bibinfo{pages}{063512} (\bibinfo{year}{2012}), \eprint{1111.2997}.

\bibitem[{\citenamefont{Bruni et~al.}(2012)\citenamefont{Bruni, Crittenden,
  Koyama, Maartens, Pitrou et~al.}}]{Bruni:2011ta}
\bibinfo{author}{\bibfnamefont{M.}~\bibnamefont{Bruni}},
  \bibinfo{author}{\bibfnamefont{R.}~\bibnamefont{Crittenden}},
  \bibinfo{author}{\bibfnamefont{K.}~\bibnamefont{Koyama}},
  \bibinfo{author}{\bibfnamefont{R.}~\bibnamefont{Maartens}},
  \bibinfo{author}{\bibfnamefont{C.}~\bibnamefont{Pitrou}},
  \bibnamefont{et~al.}, \bibinfo{journal}{Phys.Rev.}
  \textbf{\bibinfo{volume}{D85}}, \bibinfo{pages}{041301}
  (\bibinfo{year}{2012}), \eprint{1106.3999}.

\bibitem[{\citenamefont{Bartolo et~al.}(2005)\citenamefont{Bartolo, Matarrese,
  and Riotto}}]{Bar05}
\bibinfo{author}{\bibfnamefont{N.}~\bibnamefont{Bartolo}},
  \bibinfo{author}{\bibfnamefont{S.}~\bibnamefont{Matarrese}},
  \bibnamefont{and} \bibinfo{author}{\bibfnamefont{A.}~\bibnamefont{Riotto}},
  \bibinfo{journal}{JCAP} \textbf{\bibinfo{volume}{0510}}, \bibinfo{pages}{010}
  (\bibinfo{year}{2005}), \eprint{astro-ph/0501614}.

\bibitem[{\citenamefont{{Bartolo} et~al.}(2010)\citenamefont{{Bartolo},
  {Matarrese}, {Pantano}, and {Riotto}}}]{Bar10}
\bibinfo{author}{\bibfnamefont{N.}~\bibnamefont{{Bartolo}}},
  \bibinfo{author}{\bibfnamefont{S.}~\bibnamefont{{Matarrese}}},
  \bibinfo{author}{\bibfnamefont{O.}~\bibnamefont{{Pantano}}},
  \bibnamefont{and} \bibinfo{author}{\bibfnamefont{A.}~\bibnamefont{{Riotto}}},
  \bibinfo{journal}{Classical and Quantum Gravity}
  \textbf{\bibinfo{volume}{27}}, \bibinfo{pages}{124009}
  (\bibinfo{year}{2010}), \eprint{1002.3759}.

\bibitem[{\citenamefont{Bruni et~al.}(2014{\natexlab{b}})\citenamefont{Bruni,
  Hidalgo, Meures, and Wands}}]{Bruni:2013qta}
\bibinfo{author}{\bibfnamefont{M.}~\bibnamefont{Bruni}},
  \bibinfo{author}{\bibfnamefont{J.~C.} \bibnamefont{Hidalgo}},
  \bibinfo{author}{\bibfnamefont{N.}~\bibnamefont{Meures}}, \bibnamefont{and}
  \bibinfo{author}{\bibfnamefont{D.}~\bibnamefont{Wands}},
  \bibinfo{journal}{Astrophys.J.} \textbf{\bibinfo{volume}{785}},
  \bibinfo{pages}{2} (\bibinfo{year}{2014}{\natexlab{b}}), \eprint{1307.1478}.

\bibitem[{\citenamefont{Bruni et~al.}(2014{\natexlab{c}})\citenamefont{Bruni,
  Hidalgo, and Wands}}]{Bruni:2014xma}
\bibinfo{author}{\bibfnamefont{M.}~\bibnamefont{Bruni}},
  \bibinfo{author}{\bibfnamefont{J.~C.} \bibnamefont{Hidalgo}},
  \bibnamefont{and} \bibinfo{author}{\bibfnamefont{D.}~\bibnamefont{Wands}},
  \bibinfo{journal}{Astrophys.J.} \textbf{\bibinfo{volume}{794}},
  \bibinfo{pages}{L11} (\bibinfo{year}{2014}{\natexlab{c}}),
  \eprint{1405.7006}.

\bibitem[{\citenamefont{Kopp et~al.}(2014)\citenamefont{Kopp, Uhlemann, and
  Haugg}}]{Kopp:2013tqa}
\bibinfo{author}{\bibfnamefont{M.}~\bibnamefont{Kopp}},
  \bibinfo{author}{\bibfnamefont{C.}~\bibnamefont{Uhlemann}}, \bibnamefont{and}
  \bibinfo{author}{\bibfnamefont{T.}~\bibnamefont{Haugg}},
  \bibinfo{journal}{JCAP} \textbf{\bibinfo{volume}{1403}}, \bibinfo{pages}{018}
  (\bibinfo{year}{2014}), \eprint{1312.3638}.

\bibitem[{\citenamefont{Adamek et~al.}(2014)\citenamefont{Adamek, Durrer, and
  Kunz}}]{Adamek:2014xba}
\bibinfo{author}{\bibfnamefont{J.}~\bibnamefont{Adamek}},
  \bibinfo{author}{\bibfnamefont{R.}~\bibnamefont{Durrer}}, \bibnamefont{and}
  \bibinfo{author}{\bibfnamefont{M.}~\bibnamefont{Kunz}},
  \bibinfo{journal}{Class.Quant.Grav.} \textbf{\bibinfo{volume}{31}},
  \bibinfo{pages}{234006} (\bibinfo{year}{2014}), \eprint{1408.3352}.

\bibitem[{\citenamefont{Green and Wald}(2014)}]{Green:2014aga}
\bibinfo{author}{\bibfnamefont{S.~R.} \bibnamefont{Green}} \bibnamefont{and}
  \bibinfo{author}{\bibfnamefont{R.~M.} \bibnamefont{Wald}},
  \bibinfo{journal}{Class.Quant.Grav.} \textbf{\bibinfo{volume}{31}},
  \bibinfo{pages}{234003} (\bibinfo{year}{2014}), \eprint{1407.8084}.

\bibitem[{\citenamefont{{Villa} et~al.}(2014)\citenamefont{{Villa}, {Verde},
  and {Matarrese}}}]{2014CQGra..31w4005V}
\bibinfo{author}{\bibfnamefont{E.}~\bibnamefont{{Villa}}},
  \bibinfo{author}{\bibfnamefont{L.}~\bibnamefont{{Verde}}}, \bibnamefont{and}
  \bibinfo{author}{\bibfnamefont{S.}~\bibnamefont{{Matarrese}}},
  \bibinfo{journal}{Classical and Quantum Gravity}
  \textbf{\bibinfo{volume}{31}}, \bibinfo{eid}{234005} (\bibinfo{year}{2014}),
  \eprint{1409.4738}.

\bibitem[{\citenamefont{Rampf et~al.}(2014)\citenamefont{Rampf, Rigopoulos, and
  Valkenburg}}]{Rigopoulos:2014rqa}
\bibinfo{author}{\bibfnamefont{C.}~\bibnamefont{Rampf}},
  \bibinfo{author}{\bibfnamefont{G.}~\bibnamefont{Rigopoulos}},
  \bibnamefont{and}
  \bibinfo{author}{\bibfnamefont{W.}~\bibnamefont{Valkenburg}},
  \bibinfo{journal}{Class.Quant.Grav.} \textbf{\bibinfo{volume}{31}},
  \bibinfo{pages}{234004} (\bibinfo{year}{2014}), \eprint{1409.6549}.

\bibitem[{\citenamefont{Yoo}(2014)}]{Yoo:2014kpa}
\bibinfo{author}{\bibfnamefont{J.}~\bibnamefont{Yoo}},
  \bibinfo{journal}{Class.Quant.Grav.} \textbf{\bibinfo{volume}{31}},
  \bibinfo{pages}{234001} (\bibinfo{year}{2014}), \eprint{1409.3223}.

\bibitem[{\citenamefont{{Peebles}}(1980)}]{Peebles:1980za}
\bibinfo{author}{\bibfnamefont{P.~J.~E.} \bibnamefont{{Peebles}}},
  \emph{\bibinfo{title}{{The large-scale structure of the universe}}}
  (\bibinfo{year}{1980}).

\bibitem[{\citenamefont{{Peacock}}(1999)}]{Peacockbook}
\bibinfo{author}{\bibfnamefont{J.~A.} \bibnamefont{{Peacock}}},
  \emph{\bibinfo{title}{{Cosmological Physics}}} (\bibinfo{year}{1999}).

\bibitem[{\citenamefont{Ma and Bertschinger}(1994)}]{Ma:1995ey}
\bibinfo{author}{\bibfnamefont{C.-P.} \bibnamefont{Ma}} \bibnamefont{and}
  \bibinfo{author}{\bibfnamefont{E.}~\bibnamefont{Bertschinger}},
  \bibinfo{journal}{Astrophys. J.}  (\bibinfo{year}{1994}),
  \eprint{astro-ph/9401007}.

\bibitem[{\citenamefont{Weinberg}(1972)}]{Weinberg:1972zz}
\bibinfo{author}{\bibfnamefont{S.}~\bibnamefont{Weinberg}},
  \emph{\bibinfo{title}{{Gravitation and Cosmology: Principles and Applications
  of the General Theory of Relativity}}} (\bibinfo{year}{1972}),
  \bibinfo{note}{wiley \& Sons, New York}.

\bibitem[{\citenamefont{{Poisson} and {Will}}(2014)}]{2014grav.book.....P}
\bibinfo{author}{\bibfnamefont{E.}~\bibnamefont{{Poisson}}} \bibnamefont{and}
  \bibinfo{author}{\bibfnamefont{C.~M.} \bibnamefont{{Will}}},
  \emph{\bibinfo{title}{{Gravity}}} (\bibinfo{publisher}{Cambridge, UK:
  Cambridge University Press}, \bibinfo{year}{2014}).

\bibitem[{\citenamefont{Futamase and Schutz}(1983)}]{PhysRevD.28.2363}
\bibinfo{author}{\bibfnamefont{T.}~\bibnamefont{Futamase}} \bibnamefont{and}
  \bibinfo{author}{\bibfnamefont{B.~F.} \bibnamefont{Schutz}},
  \bibinfo{journal}{Phys. Rev. D} \textbf{\bibinfo{volume}{28}},
  \bibinfo{pages}{2363} (\bibinfo{year}{1983}).

\bibitem[{\citenamefont{Chandresekhar}(1965)}]{Chandrasekhar:1965ch}
\bibinfo{author}{\bibfnamefont{S.}~\bibnamefont{Chandresekhar}},
  \bibinfo{journal}{Astrophysical J.} \textbf{\bibinfo{volume}{142}},
  \bibinfo{pages}{1488} (\bibinfo{year}{1965}).

\bibitem[{\citenamefont{Stachel}(2006)}]{stachel}
\bibinfo{author}{\bibfnamefont{J.}~\bibnamefont{Stachel}}, in
  \emph{\bibinfo{booktitle}{Topics in Mathematical Physics, General Relativity
  and Cosmology}} (\bibinfo{year}{2006}), vol.~\bibinfo{volume}{1}, p.
  \bibinfo{pages}{453}.

\bibitem[{\citenamefont{Misner et~al.}(1973)\citenamefont{Misner, Thorne, and
  Wheeler}}]{Misner:1973mi}
\bibinfo{author}{\bibfnamefont{C.}~\bibnamefont{Misner}},
  \bibinfo{author}{\bibfnamefont{K.}~\bibnamefont{Thorne}}, \bibnamefont{and}
  \bibinfo{author}{\bibfnamefont{J.}~\bibnamefont{Wheeler}},
  \bibinfo{journal}{\itshape{Gravitation}\normalfont, Freeman, San Francisco}
  (\bibinfo{year}{1973}).

\bibitem[{\citenamefont{Futamase}(1988)}]{PhysRevLett.61.2175}
\bibinfo{author}{\bibfnamefont{T.}~\bibnamefont{Futamase}},
  \bibinfo{journal}{Phys. Rev. Lett.} \textbf{\bibinfo{volume}{61}},
  \bibinfo{pages}{2175} (\bibinfo{year}{1988}).

\bibitem[{\citenamefont{Tomita}(1988)}]{Tomita:1988to}
\bibinfo{author}{\bibfnamefont{K.}~\bibnamefont{Tomita}},
  \bibinfo{journal}{Prog.Theor.Phys.} \textbf{\bibinfo{volume}{79}},
  \bibinfo{pages}{258} (\bibinfo{year}{1988}).

\bibitem[{\citenamefont{Tomita}(1991)}]{Tomita:1991to}
\bibinfo{author}{\bibfnamefont{K.}~\bibnamefont{Tomita}},
  \bibinfo{journal}{Prog.Theor.Phys.} \textbf{\bibinfo{volume}{85}},
  \bibinfo{pages}{1041} (\bibinfo{year}{1991}).

\bibitem[{\citenamefont{Futamase}(1996)}]{PhysRevD.53.681}
\bibinfo{author}{\bibfnamefont{T.}~\bibnamefont{Futamase}},
  \bibinfo{journal}{Phys. Rev. D} \textbf{\bibinfo{volume}{53}},
  \bibinfo{pages}{681} (\bibinfo{year}{1996}).

\bibitem[{\citenamefont{Kofman and Pogosian}(1995)}]{Kofman:1994kf}
\bibinfo{author}{\bibfnamefont{L.}~\bibnamefont{Kofman}} \bibnamefont{and}
  \bibinfo{author}{\bibfnamefont{D.}~\bibnamefont{Pogosian}},
  \bibinfo{journal}{Astrophys.J.} \textbf{\bibinfo{volume}{442}},
  \bibinfo{pages}{30} (\bibinfo{year}{1995}), \eprint{astro-ph/9403029}.

\bibitem[{\citenamefont{Takada and Futamase}(1999)}]{Takada:1997bk}
\bibinfo{author}{\bibfnamefont{M.}~\bibnamefont{Takada}} \bibnamefont{and}
  \bibinfo{author}{\bibfnamefont{T.}~\bibnamefont{Futamase}},
  \bibinfo{journal}{{ Mon. Not. Roy. Astron. Soc.}}
  \textbf{\bibinfo{volume}{306,1}}, \bibinfo{pages}{64} (\bibinfo{year}{1999}),
  \eprint{astro-ph/9711344}.

\bibitem[{\citenamefont{Matarrese and Terranova}(1996)}]{Matarrese:1995sb}
\bibinfo{author}{\bibfnamefont{S.}~\bibnamefont{Matarrese}} \bibnamefont{and}
  \bibinfo{author}{\bibfnamefont{D.}~\bibnamefont{Terranova}},
  \bibinfo{journal}{Mon. Not. Roy. Astron. Soc.}
  \textbf{\bibinfo{volume}{283}}, \bibinfo{pages}{400} (\bibinfo{year}{1996}),
  \eprint{astro-ph/9511093}.

\bibitem[{\citenamefont{Hwang et~al.}(2008)\citenamefont{Hwang, Noh, and
  Puetzfeld}}]{Hwang:2005mg}
\bibinfo{author}{\bibfnamefont{J.-C.} \bibnamefont{Hwang}},
  \bibinfo{author}{\bibfnamefont{H.}~\bibnamefont{Noh}}, \bibnamefont{and}
  \bibinfo{author}{\bibfnamefont{D.}~\bibnamefont{Puetzfeld}},
  \bibinfo{journal}{JCAP} \textbf{\bibinfo{volume}{0803}}, \bibinfo{pages}{010}
  (\bibinfo{year}{2008}), \eprint{astro-ph/0507085}.

\bibitem[{\citenamefont{Szekeres and Rainsford}(2000)}]{Szekeres:1999ny}
\bibinfo{author}{\bibfnamefont{P.}~\bibnamefont{Szekeres}} \bibnamefont{and}
  \bibinfo{author}{\bibfnamefont{T.}~\bibnamefont{Rainsford}},
  \bibinfo{journal}{Gen. Rel. Grav.} \textbf{\bibinfo{volume}{32}},
  \bibinfo{pages}{479} (\bibinfo{year}{2000}), \eprint{gr-qc/9903056}.

\bibitem[{\citenamefont{Szekeres}(2000)}]{Szekeres:2000ki}
\bibinfo{author}{\bibfnamefont{P.}~\bibnamefont{Szekeres}},
  \bibinfo{journal}{Gen. Rel. Grav.} \textbf{\bibinfo{volume}{32}},
  \bibinfo{pages}{1025} (\bibinfo{year}{2000}).

\bibitem[{\citenamefont{Carbone and Matarrese}(2005)}]{Carbone:2004iv}
\bibinfo{author}{\bibfnamefont{C.}~\bibnamefont{Carbone}} \bibnamefont{and}
  \bibinfo{author}{\bibfnamefont{S.}~\bibnamefont{Matarrese}},
  \bibinfo{journal}{Phys. Rev.} \textbf{\bibinfo{volume}{D71}},
  \bibinfo{pages}{043508} (\bibinfo{year}{2005}), \eprint{astro-ph/0407611}.

\bibitem[{\citenamefont{{Matarrese} et~al.}(1998)\citenamefont{{Matarrese},
  {Mollerach}, and {Bruni}}}]{poisgaug}
\bibinfo{author}{\bibfnamefont{S.}~\bibnamefont{{Matarrese}}},
  \bibinfo{author}{\bibfnamefont{S.}~\bibnamefont{{Mollerach}}},
  \bibnamefont{and} \bibinfo{author}{\bibfnamefont{M.}~\bibnamefont{{Bruni}}},
  \bibinfo{journal}{\prd} \textbf{\bibinfo{volume}{58}}, \bibinfo{eid}{043504}
  (\bibinfo{year}{1998}), \eprint{arXiv:astro-ph/9707278}.

\bibitem[{\citenamefont{{Rampf}}(2014)}]{2014PhRvD..89f3509R}
\bibinfo{author}{\bibfnamefont{C.}~\bibnamefont{{Rampf}}},
  \bibinfo{journal}{\prd} \textbf{\bibinfo{volume}{89}}, \bibinfo{eid}{063509}
  (\bibinfo{year}{2014}), \eprint{1307.1725}.

\bibitem[{\citenamefont{{Rampf} and {Wiegand}}(2014)}]{2014PhRvD..90l3503R}
\bibinfo{author}{\bibfnamefont{C.}~\bibnamefont{{Rampf}}} \bibnamefont{and}
  \bibinfo{author}{\bibfnamefont{A.}~\bibnamefont{{Wiegand}}},
  \bibinfo{journal}{\prd} \textbf{\bibinfo{volume}{90}}, \bibinfo{eid}{123503}
  (\bibinfo{year}{2014}), \eprint{1409.2688}.

\bibitem[{\citenamefont{Blanchet}(2006)}]{lrr-2006-4}
\bibinfo{author}{\bibfnamefont{L.}~\bibnamefont{Blanchet}},
  \bibinfo{journal}{Living Rev. Rel.} \textbf{\bibinfo{volume}{9}}
  (\bibinfo{year}{2006}),
  \urlprefix\url{http://www.livingreviews.org/lrr-2006-4}.

\bibitem[{\citenamefont{Will}(1987)}]{test}
\bibinfo{author}{\bibfnamefont{C.}~\bibnamefont{Will}},
  \emph{\bibinfo{title}{in: 300 Years of Gravitation}}
  (\bibinfo{publisher}{Cambridge, Cambridge University Press},
  \bibinfo{year}{1987}), \bibinfo{note}{ed. S. Hawking and W. Israel}.

\bibitem[{\citenamefont{Bertschinger}(2000)}]{mit-notes}
\bibinfo{author}{\bibfnamefont{E.}~\bibnamefont{Bertschinger}},
  \bibinfo{journal}{M.I.T. General Relativity Notes}
  \textbf{\bibinfo{volume}{6}} (\bibinfo{year}{2000}),
  \eprint{http://web.mit.edu/edbert/GR/gr6.pdf}.

\bibitem[{\citenamefont{Uggla and Wainwright}(2013)}]{Uggla:2012gg}
\bibinfo{author}{\bibfnamefont{C.}~\bibnamefont{Uggla}} \bibnamefont{and}
  \bibinfo{author}{\bibfnamefont{J.}~\bibnamefont{Wainwright}},
  \bibinfo{journal}{Gen.Rel.Grav.} \textbf{\bibinfo{volume}{45}},
  \bibinfo{pages}{643} (\bibinfo{year}{2013}), \eprint{1203.4790}.

\bibitem[{\citenamefont{Thomas et~al.}(2014)\citenamefont{Thomas, Bruni, and
  Wands}}]{Thomas:2014aga}
\bibinfo{author}{\bibfnamefont{D.~B.} \bibnamefont{Thomas}},
  \bibinfo{author}{\bibfnamefont{M.}~\bibnamefont{Bruni}}, \bibnamefont{and}
  \bibinfo{author}{\bibfnamefont{D.}~\bibnamefont{Wands}}
  (\bibinfo{year}{2014}), \eprint{1403.4947}.

\bibitem[{\citenamefont{Lorentz et~al.}(1952)\citenamefont{Lorentz, Einstein,
  and Minkowski}}]{Einstein}
\bibinfo{author}{\bibfnamefont{H.}~\bibnamefont{Lorentz}},
  \bibinfo{author}{\bibfnamefont{A.}~\bibnamefont{Einstein}}, \bibnamefont{and}
  \bibinfo{author}{\bibfnamefont{H.}~\bibnamefont{Minkowski}},
  \emph{\bibinfo{title}{The principle of relativity: a collection of original
  memoirs on the special and general theory of relativity}}
  (\bibinfo{publisher}{Dover Publications}, \bibinfo{year}{1952}).

\bibitem[{\citenamefont{Pueblas and Scoccimarro}(2009)}]{Pueblas:2008uv}
\bibinfo{author}{\bibfnamefont{S.}~\bibnamefont{Pueblas}} \bibnamefont{and}
  \bibinfo{author}{\bibfnamefont{R.}~\bibnamefont{Scoccimarro}},
  \bibinfo{journal}{Phys.Rev.} \textbf{\bibinfo{volume}{D80}},
  \bibinfo{pages}{043504} (\bibinfo{year}{2009}), \eprint{0809.4606}.

\bibitem[{\citenamefont{Ehlers and Buchert}(2009)}]{Ehlers:2009uv}
\bibinfo{author}{\bibfnamefont{J.}~\bibnamefont{Ehlers}} \bibnamefont{and}
  \bibinfo{author}{\bibfnamefont{T.}~\bibnamefont{Buchert}},
  \bibinfo{journal}{Gen.Rel.Grav.} \textbf{\bibinfo{volume}{41}},
  \bibinfo{pages}{2153} (\bibinfo{year}{2009}), \eprint{0907.2645}.

\bibitem[{\citenamefont{Bardeen}(1980)}]{Bardeen:1980kt}
\bibinfo{author}{\bibfnamefont{J.~M.} \bibnamefont{Bardeen}},
  \bibinfo{journal}{Phys. Rev.} \textbf{\bibinfo{volume}{D22}},
  \bibinfo{pages}{1882} (\bibinfo{year}{1980}).

\bibitem[{\citenamefont{Bruni et~al.}(1992)\citenamefont{Bruni, Dunsby, and
  Ellis}}]{Bruni:1992dg}
\bibinfo{author}{\bibfnamefont{M.}~\bibnamefont{Bruni}},
  \bibinfo{author}{\bibfnamefont{P.~K.~S.} \bibnamefont{Dunsby}},
  \bibnamefont{and} \bibinfo{author}{\bibfnamefont{G.~F.~R.}
  \bibnamefont{Ellis}}, \bibinfo{journal}{Astrophys. J.}
  \textbf{\bibinfo{volume}{395}}, \bibinfo{pages}{34} (\bibinfo{year}{1992}).

\bibitem[{\citenamefont{{Ellis, George F. R.}}({2009})}]{ellis1971Varenna}
\bibinfo{author}{\bibnamefont{{Ellis, George F. R.}}},
  \bibinfo{journal}{{General Relativity and Gravitation}}
  \textbf{\bibinfo{volume}{41}}, \bibinfo{pages}{581} (\bibinfo{year}{{2009}}).

\bibitem[{\citenamefont{Kodama and Sasaki}(1984)}]{Kodama:1985bj}
\bibinfo{author}{\bibfnamefont{H.}~\bibnamefont{Kodama}} \bibnamefont{and}
  \bibinfo{author}{\bibfnamefont{M.}~\bibnamefont{Sasaki}},
  \bibinfo{journal}{Prog. Theor. Phys. Suppl.} \textbf{\bibinfo{volume}{78}},
  \bibinfo{pages}{1} (\bibinfo{year}{1984}).

\bibitem[{\citenamefont{Bruni et~al.}(1997)\citenamefont{Bruni, Matarrese,
  Mollerach, and Sonego}}]{Bruni:1996im}
\bibinfo{author}{\bibfnamefont{M.}~\bibnamefont{Bruni}},
  \bibinfo{author}{\bibfnamefont{S.}~\bibnamefont{Matarrese}},
  \bibinfo{author}{\bibfnamefont{S.}~\bibnamefont{Mollerach}},
  \bibnamefont{and} \bibinfo{author}{\bibfnamefont{S.}~\bibnamefont{Sonego}},
  \bibinfo{journal}{Class.Quant.Grav.} \textbf{\bibinfo{volume}{14}},
  \bibinfo{pages}{2585} (\bibinfo{year}{1997}), \eprint{gr-qc/9609040}.

\bibitem[{\citenamefont{Ananda et~al.}(2007)\citenamefont{Ananda, Clarkson, and
  Wands}}]{Ananda:2006af}
\bibinfo{author}{\bibfnamefont{K.~N.} \bibnamefont{Ananda}},
  \bibinfo{author}{\bibfnamefont{C.}~\bibnamefont{Clarkson}}, \bibnamefont{and}
  \bibinfo{author}{\bibfnamefont{D.}~\bibnamefont{Wands}},
  \bibinfo{journal}{Phys.Rev.} \textbf{\bibinfo{volume}{D75}},
  \bibinfo{pages}{123518} (\bibinfo{year}{2007}), \eprint{gr-qc/0612013}.

\bibitem[{\citenamefont{Bertacca et~al.}(2015)\citenamefont{Bertacca, Bartolo,
  Bruni, Koyama, Maartens et~al.}}]{Bertacca:2015mca}
\bibinfo{author}{\bibfnamefont{D.}~\bibnamefont{Bertacca}},
  \bibinfo{author}{\bibfnamefont{N.}~\bibnamefont{Bartolo}},
  \bibinfo{author}{\bibfnamefont{M.}~\bibnamefont{Bruni}},
  \bibinfo{author}{\bibfnamefont{K.}~\bibnamefont{Koyama}},
  \bibinfo{author}{\bibfnamefont{R.}~\bibnamefont{Maartens}},
  \bibnamefont{et~al.} (\bibinfo{year}{2015}), \eprint{1501.03163}.

\bibitem[{\citenamefont{Llinares and Mota}(2014)}]{Llinares:2013jua}
\bibinfo{author}{\bibfnamefont{C.}~\bibnamefont{Llinares}} \bibnamefont{and}
  \bibinfo{author}{\bibfnamefont{D.~F.} \bibnamefont{Mota}},
  \bibinfo{journal}{Phys.Rev.} \textbf{\bibinfo{volume}{D89}},
  \bibinfo{pages}{084023} (\bibinfo{year}{2014}), \eprint{1312.6016}.

\bibitem[{\citenamefont{Milillo}(2010)}]{milillo:2010}
\bibinfo{author}{\bibfnamefont{I.}~\bibnamefont{Milillo}}, Ph.D. thesis,
  \bibinfo{school}{University of Portsmouth} (\bibinfo{year}{2010}).

\end{thebibliography}

\end{document}